\documentclass[english]{article}
\usepackage{lmodern}

\usepackage[T1]{fontenc}
\usepackage[a4paper]{geometry}
\usepackage{dsfont}
\usepackage{color}
\usepackage{babel}
\usepackage{amsmath}
\usepackage{amsthm}
\usepackage{blindtext}
\usepackage{tikz-cd}
\usepackage{bm}
\usepackage{amsmath} 
\usepackage{enumitem}
\usepackage{booktabs}
\usepackage{multirow}
\usepackage{float}
\usepackage{subfig}
\usepackage{graphicx}
\usepackage{amsthm,mathrsfs}
\usepackage{pdfpages}
\usepackage{lscape}
\usepackage{float}
\usepackage[linesnumbered,ruled,lined]{algorithm2e}
\usepackage{makecell}
\usepackage{amssymb}
\usepackage{stmaryrd}
\usepackage{graphicx}
\usepackage{setspace}
\usepackage{esint}
\usepackage{caption}

\usepackage[authoryear]{natbib}
\usepackage{float}
\usepackage{subfig}
\usepackage{nicefrac}

\usepackage{epstopdf}
\usepackage{url} 
\usepackage{caption}
\usepackage{multirow}
\usepackage{amssymb}
\usepackage{multicol}
\usepackage{booktabs}
\usepackage{tikz}
\usepackage{bbm}
\usepackage{bm}
\usetikzlibrary{positioning, bayesnet}
\usepackage{algorithm2e}
\RestyleAlgo{ruled}

\newtheorem{theorem}{Theorem}
\newtheorem{example}{Example}
\newtheorem{definition}{Definition}
\newtheorem{proposition}{Proposition}
\newtheorem{assumption}{Assumption}
\newtheorem{assumption*}{Assumption}[section]

\usepackage[textsize=tiny]{todonotes}

\newcommand{\hatbasisk}{\hat{\mathbf{Z}}_\gamma(\mathbf{X}_{\hat{ca}_G(k)})}

\newcommand\relph[1][=]{\mathrel{\phantom{#1}}}

\def\ve{\varepsilon}
\def\bve{\boldsymbol\varepsilon}
\def\ind{\perp\!\!\!\perp}
\def\nind{\,\not\!\perp\!\!\!\perp}

\def\bB{\mathbf{B}}
\def\bM{\mathbf{M}}
\def\bR{\mathbf{R}}
\def\bv{\mathbf{v}}
\def\bbm{\mathbf{m}}
\def\bC{\mathbf{C}}
\def\bO{\mathbf{O}}
\def\bI{\mathbf{I}}
\def\bW{\mathbf{W}}
\def\bbeta{\bm{\beta}}
\def\bphi{\bm{\phi}}
\def\btheta{\bm{\theta}}
\def\bOmega{\bm{\Omega}}
\def\bA{\mathbf{A}}
\def\bU{\mathbf{U}}
\def\bC{\mathbf{C}}
\def\bD{\mathbf{D}}
\def\bG{\mathbf{G}}
\def\bF{\mathbf{F}}

\DeclareMathOperator\pa{pa}
\DeclareMathOperator\an{an}
\DeclareMathOperator\intr{in}
\DeclareMathOperator\me{me}
\DeclareMathOperator\nm{nm}
\DeclareMathOperator\leaf{leaf}
\DeclareMathOperator\iv{iv}
\DeclareMathOperator\ca{ca}
\DeclareMathOperator\cov{cov}
\DeclareMathOperator\var{var}
\DeclareMathOperator\diag{diag}
\DeclareMathOperator*\argmin{\arg\min}
\DeclareMathOperator*\argmax{\arg\max}
\DeclareMathOperator\TP{TP}
\DeclareMathOperator\RE{RE}
\DeclareMathOperator\FP{FP}
\DeclareMathOperator\FN{FN}
\DeclareMathOperator\FDP{FDP}
\DeclareMathOperator\FDR{FDR}
\DeclareMathOperator\TPR{TPR}
\DeclareMathOperator\SHD{SHD}
\DeclareMathOperator\JI{JI}

\SetKw{Return}{Return:}

\makeatletter
\newcommand{\myfnsymbol}[1]{%
  \expandafter\@myfnsymbol\csname c@#1\endcsname
}
\newcommand{\@myfnsymbol}[1]{%
  \ifcase #1
  \or 1
  \or 2
  \or 3
  \or 4
  \or \TextOrMath{\textasteriskcentered}{*}
  \or \TextOrMath{\textdagger}{\dagger}
  \fi
}
\newcommand{\affiliationA}{\@myfnsymbol{1}}
\newcommand{\affiliationB}{\@myfnsymbol{2}}
\newcommand{\affiliationC}{\@myfnsymbol{3}}
\newcommand{\affiliationD}{\@myfnsymbol{4}}
\newcommand{\equalcontribution}{\@myfnsymbol{5}}
\newcommand{\correspondingA}{\@myfnsymbol{6}}
\makeatother

\makeatother

\begin{document}
	\title{Semiparametric Causal Discovery and Inference with Invalid Instruments}
	\date{}
\author{
    Jing Zou\textsuperscript{\affiliationA},
    Wei Li\textsuperscript{\affiliationB},
    Wei Lin\textsuperscript{\affiliationC}
}
\renewcommand{\thefootnote}{\myfnsymbol{footnote}}
\maketitle
\footnotetext[1]{School of Mathematical Sciences, Peking University,
   Beijing, 100871, China. Email: jingzou@stu.pku.edu.cn}%
\footnotetext[2]{Center for Applied Statistics and School of Statistics, Renmin University of China,
Beijing, 100872, China. Email: weilistat@ruc.edu.cn}%
\footnotetext[3]{School of Mathematical Sciences and Center for Statistical Science, Peking University, 
   Beijing, 100871, China. Email: weilin@math.pku.edu.cn}

    \begin{abstract}
        Learning causal relationships among a set of variables, as encoded by a directed acyclic graph, from observational data is complicated by the presence of unobserved confounders. Instrumental variables (IVs) are a popular remedy for this issue, but most existing methods either assume the validity of all IVs or postulate a specific form of relationship, such as a linear model, between the primary variables and the IVs. To overcome these limitations, we introduce a partially linear structural equation model for causal discovery and inference that accommodates potentially invalid IVs and allows for general dependence of the primary variables on the IVs. We establish identification under this semiparametric model by constructing surrogate valid IVs, and develop a finite-sample procedure for estimating the causal structures and effects. Theoretically, we show that our procedure consistently learns the causal structures, yields asymptotically normal estimates, and effectively controls the false discovery rate in edge recovery. Simulation studies demonstrate the superiority of our method over existing competitors, and an application to inferring gene regulatory networks in Alzheimer's disease illustrates its usefulness.
        \end{abstract}
        

        \section{Introduction}
        Identifying and inferring causal relationships among a set of variables, as represented by a directed acyclic graph (DAG), is a fundamental problem in statistics and machine learning and finds applications in various fields such as systems biology \citep{triantafillou2017predicting}, psychology \citep{grosz2020taboo}, medical imaging \citep{castro2020causality}, and philosophy \citep{malinsky2018causal}. In many such cases, randomized experiments for traditional causal inference can be prohibitively expensive, time-consuming, or even impossible. As a result, learning causal relationships from observational data, known as causal discovery, has gained popularity and become an active area of research; see \citet{heinze-deml2018causal}, \citet{glymour2019review}, and \citet{vowels2022d} for comprehensive reviews. The potential existence of unobserved confounders, however, may cause violations of the Markov property in the DAG and lead to biased estimates of causal effects \citep{pearl2009causality}, posing challenges to robust causal discovery. The existing literature offers some approaches to tackling this challenge. One way is to generate less informative discoveries, such as constructing a partial ancestor graph instead of a DAG \citep{colombo2012learning}. Another way is to develop effective algorithms for deconfounding, often under additional assumptions about the confounding mechanism. For example, \cite{frot2019robust} suggested a two-stage procedure for recovering the Markov equivalence class of a DAG based on the pervasive confounding assumption \citep{chandrasekaran2012latent}, and \cite{li2024nonlinear} proposed a deconfounded estimation procedure for nonlinear causal discovery under a sublinear growth condition that separates linear confounding effects from nonlinear causal relationships. To avoid such restrictions while obtaining an accurate estimate of the DAG, here we resort to the use of instrumental variables (IVs), which provides a convenient and powerful method for resolving the issue of unobserved confounding.
        
        In the classical context of inferring a treatment--outcome relationship, there has been a rich body of work on estimating causal effects using valid IVs \citep[e.g.,][]{angrist1996identification,newey2003instrumental,clarke2012instrumental,baiocchi2014instrumental}. The validity of an IV hinges on meeting three key assumptions \citep{didelez2010assumptions}: (i) (relevance) the IV is associated with the treatment; (ii) (independence) the IV is independent of the unobserved confounders; and (iii) (exclusion) the IV has no direct effect on the outcome. Owing to the untestable independence and exclusion restrictions, the candidate IVs  may be invalid, rendering the estimates biased or inconsistent. In the presence of invalid IVs, \cite{bowden2015mendelian} and \citet{kolesar2015identification} identified causal effects by assuming that the direct effects of the IVs on the outcome are asymptotically orthogonal to their effects on the treatment. Alternatively, some studies made assumptions regarding the proportion of valid IVs among the candidate set. In particular, \cite{kang2016instrumental} and \citet{windmeijer2019use} developed Lasso-based methods for selecting valid IVs and estimating causal effects under the majority rule, which requires more than half of the candidate IVs to be valid. \cite{guo2018confidence} further proposed two-stage hard thresholding with voting for constructing confidence intervals under the more general plurality rule. Recently, \cite{sun2023semiparametric} introduced a new class of G-estimators for a semiparametric structural equation model, allowing for a flexible number of valid IVs and bypassing the IV selection step.
        
        Despite the extensive developments on IV methods in causal inference, their application to causal discovery remains underexplored. Notably, \cite{oates2016estimating} formalized the notion of conditional DAGs and developed a score-based estimation method via integer linear programming, and \cite{chen2018two} proposed a two-stage penalized least squares method for estimating a large linear structural equation model; both methods require knowing a priori a unique set of valid IVs for each primary variable. IV methods for inferring the causal direction between two traits have received attention in Mendelian randomization \citep{hemani2017orienting}, with extensions to allow for invalid IVs \citep{xue2020inferring}. Among the few attempts to learn a causal graph with invalid IVs, \cite{chen2023inference} proposed a stepwise selection procedure to select valid IVs, followed by two-stage least squares and Wald tests for inference. \citet{li2023inference} and \citet{chen2024discovery} developed a peeling algorithm for estimating the ancestral relation graph and candidate IV sets, as well as likelihood-based inference procedures for edge recovery. Still, all these methods rely on the assumption of a linear structural equation model and do not account for candidate IVs whose effects on the primary variables may be nonlinear. As widely recognized in the literature, failing to exploit such nonlinearities may result in weak IVs, deteriorate the estimation performance, or distort the causal interpretation \citep{newey1990efficient,chen2020mostly,sun2023semiparametric}.
        
        In this paper, we consider causal discovery and inference in the presence of unobserved confounders using potentially invalid IVs. Unlike existing studies that postulate linear relationships between the primary variables and the IVs, we adopt a partially linear structural equation model that leaves the functional form of these relationships unspecified. Within this semiparametric framework, we establish identification of causal structures and effects under relatively mild assumptions. Specifically, we first identify the ancestral relationships and candidate IV sets by extending the peeling algorithm of \cite{chen2024discovery} to our more general setting. Building on these results, we then construct surrogate valid IVs and derive moment conditions to identify the causal effects recursively, which generalizes the identification strategy of \citet{sun2023semiparametric} to causal graphs. We further develop a finite-sample procedure for causal discovery and inference by using distance-correlation-based independence tests and the generalized method of moments. We call our method the Partially Linear Approach to Causal Instrument-based Discovery (PLACID). Theoretically, we show that PLACID consistently learns the causal structures, yields asymptotically normal estimates, and effectively controls the false discovery rate in edge recovery.
        
        The remainder of this paper is organized as follows. Section \ref{sec:model} introduces our causal graph terminology and the partially linear structural equation model. Section \ref{sec:ident} establishes the identification results for the causal graph and causal effects. Section \ref{sec:method} presents the PLACID methodology along with its theoretical guarantees. Sections \ref{sec:simul} and \ref{sec:appl} illustrate the numerical performance of our method through simulation studies and an application to an Alzheimer’s disease dataset, respectively. Section \ref{sec:disc} concludes the paper with some discussion. All proofs are relegated to the Appendix.
        
        For a $K$-dimensional random variable $\mathbf{Z}$ and an index set $\alpha\subseteq\{1,\dots,K\}$, define $\mathbf{Z}_\alpha=(Z_s:s\in\alpha)$ and $\mathbf{Z}_{-\alpha}=(Z_s:s\notin\alpha)$. Let $\mathcal{H}(\mathbf{Z})$ denote the Hilbert space of one-dimensional functions of $\mathbf{Z}$ with mean zero and finite variance, equipped with the covariance inner product. Let $|J|$ denote the cardinality of a set $J$. For a matrix ${\bA}=(A_{ij})$, let ${\bA}_{i,\cdot}$ denote its $i$th row and ${\bA}_{\cdot,j}$ its $j$th column. For a vector $\bv$, let $\bv^T$ denote its transpose, $\|\bv\|_2$ its $L_2$-norm, and $\|\bv\|_0$ its $L_0$-norm.  Let $\mathbf{1}(\cdot)$ denote the indicator function and $\bI_p$ the $p\times p$ identity matrix. Finally, let $\hat{\mathbb{E}}_n$ denote the empirical mean operator with respect to a sample of size $n$.
        
        \section{Causal graphical model}\label{sec:model}
        Consider a causal graph $G$ with $p$ endogenous primary variables $\mathbf{Y}=(Y_1,\dots,Y_p)^T$ and $q$ exogenous secondary variables $\mathbf{X}=(X_1,\dots,X_q)^T$, both having finite variance. Specifically, we denote
        \begin{equation}
            \label{definition of DAG}
            {G}=(\mathbf{X},\mathbf{Y};\mathcal{E},\mathcal{I}),
        \end{equation}
        where $\mathcal{E}=\{(i,j):Y_i\to Y_j\}$ is the set of directed edges among $\mathbf{Y}$, and $\mathcal{I}=\{(\ell,j):X_\ell\to Y_j\}$ is the set of directed edges from $\mathbf{X}$ to $\mathbf{Y}$. Note that there is no directed edge from $\mathbf{Y}$ to $\mathbf{X}$, and thus $\mathbf{X}$ can be viewed as external interventions. Based on $G$, we adopt the following terminology: (i) the parent set of $Y_j$, $\pa_G(j)=\{k:Y_k\to Y_j\}$; (ii) if there exists a directed path from $Y_k$ to $Y_j$, then $Y_j$ is a descendant of $Y_k$, $Y_k$ is an ancestor of $Y_j$, and the ancestor set of $Y_j$ is $\an_G(j)=\{k:Y_k\to\dots\to Y_j\}$; (iii) the intervention set of $Y_j$, $\intr_G(j)=\{\ell:X_\ell\to Y_j\}$; (iv) the mediator set of $Y_k$ and $Y_j$, $\me_G(k,j)=\{i:Y_k\to\dots\to Y_i\to\dots\to Y_j\}$; (v) the non-mediator set of $Y_k$ and $Y_j$, $\nm_G(k,j)=\an_G(j)\setminus(\me_G(k,j)\cup\{k\})$; (vi) $Y_k$ is an unmediated parent of $Y_j$ if $(k,j)\in\mathcal{E}$ and $\me_G(k,j)=\emptyset$; (vii) the leaf nodes of $G$, $\leaf(G)=\{j:Y_j\text{ has no descendant in }G\}$; (viii) the length of the longest directed path from $Y_k$ to $Y_j$, $l_G(k,j)$; (ix) the height of $Y_j$, $h_G(j)$, is the length of the longest directed path from $Y_j$ to a leaf node of $G$, so that if $(k,j)\in \mathcal{E}$ then $h_G(k)>h_G(j)$, and the height of any leaf node is 0. To further illustrate these definitions, we consider a simple case in Example \ref{example 1}.
        
\begin{figure}
            \centering
            \begin{tikzcd}
               &               & X_2 \arrow[d]                              & X_4 \arrow[ld] \arrow[d] &               \\
               X_1 \arrow[r] & Y_1 \arrow[r] & Y_2 \arrow[r]                              & Y_3                      & X_3 \arrow[l] \\
               &               & \bU \arrow[lu] \arrow[u] \arrow[ru] &                          &              
            \end{tikzcd}
            \caption{An example of the causal graph $G$.}
            \label{example 1-figure}
        \end{figure}
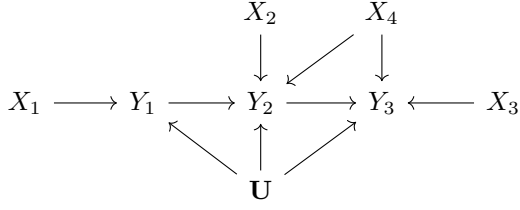

        \begin{example}
        \label{example 1}
        Consider the causal graph $G$ shown in Figure \ref{example 1-figure}, where $\bU$ represents unobserved confounders. In view of the directed path $Y_1\to Y_2\to Y_3$, the mediator set of $Y_1$ and $Y_3$ is $\me_G(1,3)=\{2\}$. The unmediated parent of $Y_3$ is $Y_2$ because $Y_2$ is the only parent of $Y_3$ with $\me_G(2,3)=\emptyset$. The height of $Y_1$ is $2$ since the longest path from $Y_1$ to a leaf node of $G$ is $Y_1\to Y_2\to Y_3$, whose length is $2$.
        \end{example}
        
        Following the idea of conditional DAGs \citep{oates2016estimating}, we wish to use $\mathbf{X}$ as candidate IVs to infer the causal relationships and effects among $\mathbf{Y}$. The secondary variables $\mathbf{X}$ are exogenous, implying that these variables satisfy the independence assumption of valid IVs. There are directed edges from $\mathbf{X}$ to $\mathbf{Y}$, but none in the opposite direction, providing the basis for the relevance assumption. Therefore, it is possible to use $\mathbf{X}$ as IVs. We further introduce our definitions of valid and candidate IVs based on causal graphs.
        
        \begin{definition}[Valid IV]
        \label{definition of valid IV}
        A secondary variable $X_\ell$ is said to be a valid IV for $Y_j$ in the causal graph $G$, if it intervenes on $Y_j$, i.e., $(\ell,j)\in \mathcal{I}$ and does not intervene on any other primary variable $Y_i$, i.e., $(\ell,i)\notin\mathcal{I}$ for all $i\ne j$. 
        \end{definition}
        
        \begin{definition}[Candidate IV]
        \label{definition of candidate IV}
        A secondary variable $X_\ell$ is said to be a candidate IV for $Y_j$ in the causal graph $G$, if it intervenes on $Y_j$, i.e., $(\ell,j)\in \mathcal{I}$ and does not intervene on any non-descendant of $Y_j$. 
        \end{definition}
        
        Accordingly, denote the set of valid IVs for $Y_j$ in $G$ by $\iv_G(j)=\{\ell:X_\ell\to Y_j,X_\ell\nrightarrow Y_i,i\ne j\}$ and the set of candidate IVs for $Y_j$ in $G$ by $\ca_G(j)=\{\ell:X_\ell\to Y_j,X_\ell\to Y_k\text{ only if }j\in\an_G(k)\}$. It is obvious that $\iv_G(j)\subseteq\ca_G(j)$. However, a candidate IV may not be valid, because it may intervene on a descendant of $Y_j$, which contradicts Definition \ref{definition of valid IV}. For example, in Figure \ref{example 1-figure}, $\iv_G(2)=\{2\}$ while $\ca_G(2)=\{2,4\}$.
        
        Since the variables $\mathbf{Y}$ are of primary interest, it is often reasonable to assume a simple relationship among $\mathbf{Y}$ while not restricting the functional forms of interactions between $\mathbf{X}$ and $\mathbf{Y}$, as proved to be useful for semiparametric modeling in econometrics \citep{engle1986semiparametric} and environmental science \citep{dominici2004improved}. Among such semiparametric models, of particular importance is the partially linear model \citep{robinson1988root}. For the causal graph $G$ in \eqref{definition of DAG}, we consider the partially linear structural equation model
        \begin{equation}
            Y_j=\sum_{i=1}^p\beta_{ij}^*Y_i+g_j(\mathbf{X}_{\intr_G(j)})+\ve_j,\quad\mathbb{E}(\ve_j)=0,\quad\mathbf{X}\ind\ve_j,\quad j=1,\dots,p.
            \label{SEM}
        \end{equation}
        Here the parameter $\beta_{ij}^*$ represents the direct causal effect of $Y_i$ on $Y_j$, where $\beta_{ij}^*\ne 0$ implies that $Y_i$ is a cause of $Y_j$, i.e., $i\in\pa_G(j)$. The function $g_j(\cdot)$ captures the causal effect of $\mathbf{X}_{\intr_G(j)}$ on $Y_j$, the form of which is unknown and not restricted to linear functions. Our interest lies in estimating the causal structures and effects among $\mathbf{Y}$, as characterized by the edge set $\mathcal{E}$ and the coefficient matrix $\bB^*=(\beta_{ij}^*)_{p\times p}$.
        
        Compared to the standard linear model, the inclusion of a nonparametric term in \eqref{SEM} enhances robustness against model misspecification \citep{florens2012instrumental,emmenegger2021regularizing}. Previous research has extensively examined the use of partially linear models in causal inference, including estimation with missing data \citep{liang2004estimation}, identifiability of partially linear structural equation models \citep{rothenhausler2018causal}, and mediation analysis \citep{hines2021robust}. However, since potential unobserved confounders have been absorbed in $\bve=(\ve_1,\dots,\ve_p)^T$, $\cov(\mathbf{Y}_{-\{j\}},\ve_j)$ may not be $\mathbf{0}$ in model \eqref{SEM}. As a result, existing estimation methods for partially linear models are unsuitable in our context, even when the causal relationships are known. Assuming that $g_j(\cdot)$ takes a linear form, \cite{chen2024discovery} proposed an insightful peeling algorithm to infer the causal relationships and effects among the primary variables $\mathbf{Y}$ using IVs. Their method, while effective in linear settings, may not be applicable to the more general semiparametric model \eqref{SEM}, because an IV deemed valid under Definition \ref{definition of valid IV} might not satisfy the criteria for a valid IV in \cite{chen2024discovery}. When no valid IVs are available, the causal parameters in their model will become unidentifiable. Specifically, \cite{chen2024discovery} defined  $X_\ell$ as a valid IV for $Y_k$ if $W_{\ell k}\ne 0$ and $W_{\ell k'}=0$ for all $k'\ne k$, where $\bW=(W_{\ell k})_{q\times p}=\mathbf{V}(\bI_p-\bB^*)$ and $\mathbf{V}$ is the linear regression coefficient of $\mathbf{Y}$ on $\mathbf{X}$, i.e., $\mathbf{V}=\{\var(\mathbf{X})\}^{-1}\cov(\mathbf{X},\mathbf{Y})$. The following example clearly demonstrates this point.
        
        \begin{example}
        \label{fail}
        Consider the causal graph $G$ shown in Figure \ref{example 1-figure}, where $Y_1=X_1^2+\ve_1$, $Y_j$ follows model \eqref{SEM} with unspecified $g_j(\cdot)$ for $j=2,3$, and $\mathbf{X}\sim N(\mathbf{0},\bI_4)$. It is easy to verify that $\mathrm{cov}(X_1,\mathbf{Y})=\mathbf{0}$, and hence $\mathbf{V}_{1,\cdot}=\mathbf{0}$ and $\bW_{1,\cdot}=\mathbf{0}$. This shows that, although $X_1$ is indeed a valid IV for $Y_1$ in $G$, it is not a valid IV for any primary variable in $\mathbf{Y}$ in the sense of \cite{chen2024discovery}. Consequently, $Y_1$ has no valid IVs, and thus the causal parameters in the model for $Y_1$ are unidentifiable.
        \end{example}
        
        In light of Example \ref{fail}, the method of \cite{chen2024discovery} fails to identify and estimate the causal parameters in model \eqref{SEM}. In contrast, our approach introduced in the next section can still ensure identification of these parameters. Furthermore, when $g_j(\cdot)$ is unknown and possibly highly nonlinear, the use of linear methods can introduce substantial estimation bias. This issue may be particularly critical for the purpose of causal discovery since it can easily lead to incorrect determination of causal directions among the primary variables. To address this problem, we next consider the discovery and estimation of DAGs using invalid IVs under model \eqref{SEM}, which has not been discussed in the existing literature.
        
        \section{Identification of the causal graph and effects}\label{sec:ident}
        In this section, we show how to identify the causal structures and effects among the primary variables $\mathbf{Y}$ using the second variables $\mathbf{X}$. We first make the following assumptions.
        
        \begin{assumption}
        \label{assumption-independent X}
        The candidate IVs are independent of each other, i.e., $X_i\ind X_j$ for all $i,j=1,\dots,q$ and $i\ne j$.
        \end{assumption}
        
        Assumption \ref{assumption-independent X} is reasonable and common in Mendelian randomization studies, where genetic variants from different gene regions are often used as IVs. To meet this assumption, one can employ well-established tools for linkage disequilibrium clumping such as PLINK \citep{purcell2007plink} to select a set of independent genetic variants. This practice is prevalent and generally accepted within the field \citep[e.g.,][]{bowden2016consistent,hartwig2017robust,zhao2020statistical,ye2021debiased}. The independence assumption has also been made in existing methods for causal discovery. For instance, \cite{neto2010causal} and \cite{ongen2017estimating} treated expression quantitative trait loci (eQTLs) as secondary variables and exploited their independence when evaluating genetic causality.
        
        \begin{assumption}
        \label{assumption-unmediated parent}
        Whenever $X_\ell$ intervenes on an unmediated parent of $Y_j$, $X_\ell\nind Y_j$.
        \end{assumption}
        
        Assumption \ref{assumption-unmediated parent} is essentially the faithfulness assumption in causal discovery \citep{spirtes2001causation,peters2017elements}. It requires that when $X_\ell$ intervenes on both $Y_j$ and its unmediated parent, the dependencies between $X_\ell$ and $Y_j$ should not be canceled out.
        
        \begin{assumption}
        \label{assumption-valid IV}
        For each primary variable $Y_j$, there are at least $\gamma\ge1$ valid IVs, i.e., $|\iv_G(j)|\ge\gamma$ for all $j=1,\dots,p$.
        \end{assumption}
        
        When there are only two primary variables and the causal direction between them is known, Assumption \ref{assumption-valid IV} is the same as the one imposed in \cite{sun2023semiparametric}. The value of $\gamma$ can be specified based on prior knowledge. In general, it is possible to set $\gamma=1$ by assuming only the existence of a valid IV for each $Y_j$, as done by \cite{li2023inference} and \cite{zilinskas2024inferring}. However, these studies did not account for unobserved confounders. In the presence of unobserved confounding, \cite{chen2024discovery} assumed the majority rule as in \cite{kang2016instrumental}, which is stronger than Assumption \ref{assumption-valid IV} since it requires at least half of the candidate IVs to be valid for each primary variable.
        
        For causal inference between an exposure and an outcome of interest under a semiparametric model similar to \eqref{SEM}, \cite{sun2023semiparametric} introduced a system of moment conditions for identification in a union of causal models where at least $\gamma$ of the candidate IVs are valid but their identities are unknown. They then proposed a class of G-estimators \citep{robins1992estimating,vansteelandt2014structural} and developed semiparametric efficiency theory. To extend this idea to our causal graph setting, we first give the following definition.
        
        \begin{definition}
        \label{definition of IH}
        For each primary variable $Y_j$ and an index set $\alpha_j\subseteq\ca_G(j)$ with $|\alpha_j|\ge\gamma$, define the subspace
        \[
        \mathcal{D}(\alpha_j)=\bigl\{d(\mathbf{X}_{\ca_G(j)}):\mathbb{E}\{d(\mathbf{X}_{\ca_G(j)})\mid\mathbf{X}_{\ca_G(j)\setminus\alpha_j}\}=0\bigr\}\cap \mathcal{H}(\mathbf{X}_{\ca_G(j)}).
        \]
        Further, define the intersection of all possible $\mathcal{D}(\alpha_j)$ by
        \[
        \mathcal{Z}_\gamma(j)=\bigcap_{\alpha_j\subseteq\ca_G(j),|\alpha_j|\ge \gamma}\mathcal{D}(\alpha_j).
        \]
        \end{definition}
        
        For discrete $\mathbf{X}_{\ca_G(j)}$, the space $\mathcal{H}(\mathbf{X}_{\ca_G(j)})$ can be spanned by a finite number of orthogonal functions. Since $\mathcal{Z}_\gamma(j)\subseteq\mathcal{H}(\mathbf{X}_{\ca_G(j)})$, we can stack all basis functions of $\mathcal{Z}_\gamma(j)$ into a vector denoted by $\mathbf{Z}_\gamma(\mathbf{X}_{\ca_G(j)})$. For continuous $\mathbf{X}_{\ca_G(j)}$, however, $\mathcal{Z}_\gamma(j)$ forms an infinite-dimensional Hilbert space. 
        We follow the general strategy in \cite{newey1993efficient} and \cite{tchetgen2010doubly} by selecting a basis set of functions $\{\phi_s(\mathbf{X}_{\ca_G(j)})\}_{s=1}^\infty$ that are dense in $\mathcal{Z}_\gamma(j)$, such as tensor products of trigonometric, wavelet, or polynomial bases. We can then construct $\mathbf{Z}_\gamma(\mathbf{X}_{\ca_G}(j))$ from a finite subset of these basis functions.
        
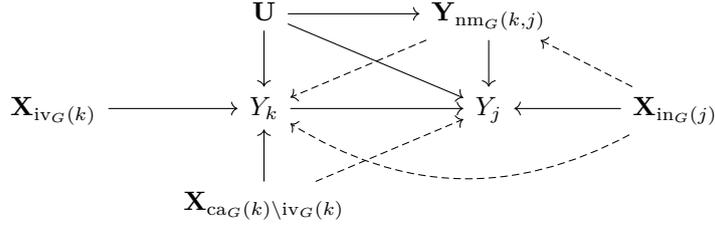
\begin{figure}
            \centering
        \begin{tikzcd}
                                       & \bU \arrow[d] \arrow[r] \arrow[rd]                                   & {\mathbf{Y}_{\nm_G(k,j)}} \arrow[ld, dashed] \arrow[d] &                                                                                 \\
        \mathbf{X}_{\iv_G(k)} \arrow[r] & Y_k \arrow[r]                                                      & Y_j                                          & \mathbf{X}_{\intr_G(j)} \arrow[l] \arrow[lu, dashed] \arrow[ll, dashed, bend left] \\
                                       & \mathbf{X}_{\ca_G(k)\setminus\iv_G(k)} \arrow[u] \arrow[ru, dashed] &                                              &                                                                                
        \end{tikzcd}
            \caption{Causal relationships in the case where $\me_G(k,j)=\emptyset$.}
            \label{fig-nonmediator}
        \end{figure}

        Below we briefly explain how $\mathbf{Z}_\gamma(\mathbf{X}_{\ca_G(j)})$ can serve as surrogate IVs for identification. Consider the simple case where $\me_G(k,j)=\emptyset$, meaning that the total causal effect of $Y_k$ on $Y_j$ is the direct effect $\beta_{kj}^*$. The causal relationships among the relevant variables for this case are illustrated in Figure \ref{fig-nonmediator}, with dashed lines indicating that the relationships may exist.
        It is easy to see that $\mathbf{Y}_{\nm_G(k,j)}\ind\mathbf{X}_{\ca_G(k)}$ and $\mathbf{X}_{\intr_G(j)}\ind\mathbf{X}_{\ca_G(k)}\mid\mathbf{X}_{\ca_G(k)\setminus \iv_G(k)}$. Then, for any $d(\mathbf{X}_{\ca_G(k)})\in\mathcal{D}(\iv_G(k))$, we have
        \begin{equation}
        \label{gmm-nonmediator}
            \begin{aligned}
            &\mathbb{E}\{d(\mathbf{X}_{\ca_G(k)})(Y_j-\beta^*_{kj}Y_k)\}\\
            &\quad=\mathbb{E}\Biggl\{d(\mathbf{X}_{\ca_G(k)})\Biggl(\sum_{i\in \nm_G(k,j)}\beta^*_{ij}Y_i+g_j(\mathbf{X}_{\intr_G(j)})+\ve_j\Biggr)\Biggr\}\\
            &\quad=\mathbb{E}\{d(\mathbf{X}_{\ca_G(k)})g_j(\mathbf{X}_{\intr_G(j)})\}\\
            &\quad=\mathbb{E}\bigl[\mathbb{E}\{d(\mathbf{X}_{\ca_G(k)})\mid\mathbf{X}_{\ca_G(k)\setminus \iv_G(k)}\} \mathbb{E}\{g_j(\mathbf{X}_{\intr_G(j)})\mid \mathbf{X}_{\ca_G(k)\setminus\iv_G(k)}\}\bigr]\\
            &\quad=0.
        \end{aligned}
        \end{equation}
        Although $\iv_G(k)$ is unknown, there must exist a subset $\alpha_k\subseteq\ca_G(k)$ as described in Definition \ref{definition of IH} such that $\alpha_k=\iv_G(k)$. By the definition of $\mathcal{Z}_\gamma(k)$, we have $\mathcal{Z}_\gamma(k)\subseteq\mathcal{D}(\iv_G(k))$. Consequently, all random variables in $\mathcal{Z}_\gamma(k)$ must satisfy the moment condition \eqref{gmm-nonmediator}, or equivalently $\mathbb{E}\{\mathbf{Z}_\gamma(\mathbf{X}_{\ca_G(k)})(Y_j-\beta^*_{kj}Y_k)\}=\mathbf{0}$ for the basis functions $\mathbf{Z}_\gamma(\mathbf{X}_{\ca_G(k)})$ of $\mathcal{Z}_\gamma(k)$. In other words, we can construct surrogate IVs $\mathbf{Z}_\gamma(\mathbf{X}_{\ca_G(k)})$ based on $\mathbf{X}_{\ca_G(k)}$ without the need to know the valid IVs.
        
        \begin{assumption}
        \label{assumption-L0-norm} 
        For each primary variable $Y_k$ with descendants, 
        \[
        \|\mathbb{E}\{\mathbf{Z}_\gamma(\mathbf{X}_{\ca_G(k)}) Y_k\}\|_0>0.
        \]
        \end{assumption}
        
        Since $\mathbb{E}\{\mathbf{Z}_\gamma(\mathbf{X}_{\ca_G(k)})\}=\mathbf{0}$ by Definition \ref{definition of IH}, Assumption \ref{assumption-L0-norm} stems from the necessity for the surrogate IVs $\mathbf{Z}_\gamma(\mathbf{X}_{\ca_G(k)})$ to satisfy the relevance assumption. Additionally, because $\mathbf{Z}_\gamma(\mathbf{X}_{\ca_G(k)})$ characterizes the function space $\mathcal{Z}_\gamma(k)$, Assumption \ref{assumption-L0-norm} entails the existence of a random variable in $\mathcal{Z}_\gamma(k)$ that is correlated with $Y_k$. This requirement is not stringent, in view of the fact that all the candidate IVs of $Y_k$ are correlated with $Y_k$. Moreover, as $\gamma$ increases, this assumption becomes milder. As an illustration, consider the case where all candidate IVs take values 0 or 1. If we believe that $m=|\ca_G(k)|$ candidate IVs of $Y_k$ are all valid, then there are $2^m-1$ basis functions in $\mathbf{Z}_\gamma(\mathbf{X}_{\ca_G(k)})$, and Assumption \ref{assumption-L0-norm} requires at least one of these basis functions to be correlated with $Y_k$.
        
        We are now ready to state our main identification result.
        
        \begin{theorem}
        \label{thm:ident}
        Suppose that Assumptions \ref{assumption-independent X}--\ref{assumption-L0-norm} hold. For the causal graph $G=(\mathbf{X},\mathbf{Y};\mathcal{E},\mathcal{I})$, the edge set $\mathcal{E}$ and the causal parameters $\{\beta^*_{ij}\}_{i\in\pa_G(j)}$ in model \eqref{SEM} are identifiable.
        \end{theorem}
        
        The proof of Theorem \ref{thm:ident} builds on the general strategy proposed by \cite{chen2024discovery}, which consists of first identifying the ancestral relation graph (ARG) and then recovering the DAG. However, rather than focusing on linear relationships between $\mathbf{X}$ and $\mathbf{Y}$, here we extend the analysis to the more complex setting of model \eqref{SEM}. We begin by identifying the ARG to roughly capture the causal directions among $\mathbf{Y}$ and obtain the candidate IV sets. We then establish the identification of the causal effects $\bB^*$ based on the identified ARG and candidate IV sets. These two steps are outlined in Sections \ref{sec:ident_graph} and \ref{sec:ident_effects}.
        
        \subsection{Identification of the ARG and candidate IV sets}\label{sec:ident_graph}
        In this subsection, we show that the ARG and candidate IV sets are identifiable based on the relationships between $\mathbf{X}$ and $\mathbf{Y}$. First, we introduce the definition of an ARG.
        
        \begin{definition}[Ancestral relation graph]
        \label{definition of ARG}
        For a causal graph ${G}=(\mathbf{X},\mathbf{Y};\mathcal{E},\mathcal{I})$, its ancestral relation graph is defined by $G^+=(\mathbf{X},\mathbf{Y};\mathcal{E}^+,\mathcal{I}^+)$, where
        \[
        \mathcal{E}^+=\{(k,j):k\in\an_G(j)\},\quad\mathcal{I}^+=\Biggl\{(\ell,j):\ell\in\bigcup_{k\in\an_G(j)\cup\{j\}}\intr_G(k)\Biggr\}.
        \]
        \end{definition}
        
        The ARG $G^+$ describes the ancestral relationships among the nodes in $G$. Specifically, if there exists a directed path from $Y_i$ to $Y_j$ in $G$, then $(i,j)\in\mathcal{E}^+$. Similarly, if there exits a directed path from $X_\ell$ to $Y_j$ in $G$, then $(\ell,j)\in\mathcal{I}^+$. To recover $\mathcal{E}^+$, we need only identify all edges originating from the unmediated parents of each node in $G$. Note that we can derive the mediator sets $\{\me_G(k,j)\}_{(k,j)\in\mathcal{E}^+}$ and the lengths $\{l_G(k,j)\}_{(k,j)\in\mathcal{E}^+}$ from $G^+$ since $\me_G(k,j)=\me_{G^+}(k,j)$ and $l_G(k,j)=l_{G^+}(k,j)$.
        
        The next two propositions generalize Propositions 1 and 2 in \citet{chen2024discovery} to the semiparametric model \eqref{SEM} and are the key ingredients for the identification of $G^+$ and candidate IV sets.
        
        \begin{proposition}
        \label{leafnodes}
        Suppose that Assumptions \ref{assumption-independent X}--\ref{assumption-valid IV} hold. Then there exists some $X_\ell$ such that $X_\ell\nind Y_k$ and $X_\ell\ind Y_{k'}$ for all $k'\ne k$ if and only if $Y_k$ is a leaf node of $G$. Furthermore, such an $X_\ell$ is a valid IV for $Y_k$ in $G$.
        \end{proposition}
        
        Proposition \ref{leafnodes} shows that the leaves of ${G}$ and their valid IVs are identifiable:
        \begin{align*}
            \leaf(G)&=\{k:\text{for some }\ell,\ X_\ell\nind Y_k\text{ and }X_\ell\ind Y_{k'}\text{ for all }k'\ne k\},\\
            \iv_G(k)&=\{\ell:X_\ell\nind Y_k\text{ and }X_\ell\ind Y_{k'}\text{ for all }k'\ne k\},\quad k\in\leaf(G).
        \end{align*}
        
        Once the leaves of ${G}$ have been identified, we can remove these nodes along with their valid IVs to obtain a subgraph $G^-=(\mathbf{X}^-,\mathbf{Y}^-;\mathcal{E}^-,\mathcal{I}^-)$, where $\mathbf{X}^-=\mathbf{X}\setminus\bigcup_{k\in\leaf(G)}\mathbf{X}_{\iv_G(k)}$, $\mathbf{Y}^-=\mathbf{Y}\setminus\mathbf{Y}_{\leaf(G)}$, and $\mathcal{E}^-$ and $\mathcal{I}^-$ denote the remaining edges from $\mathcal{E}$ and $\mathcal{I}$, respectively. By Definition \ref{definition of valid IV}, it is clear that $\iv_G(j)\subseteq\iv_{G^-}(j)$ for all $Y_j\in\mathbf{Y}^-$, implying that Assumption \ref{assumption-valid IV} holds in $G^-$. Assumptions \ref{assumption-independent X} and \ref{assumption-unmediated parent} also hold naturally in the subgraph $G^-$. Therefore, Proposition \ref{leafnodes} remains applicable to $G^-$, so that $\leaf(G^-)$ and $\iv_{G^-}(k)$ for $k\in\leaf(G^-)$ are identifiable. By iteratively applying this method to identify and remove the leaves of the current graph, a topological order among the variables in $\mathbf{Y}$ can be determined. During this process, the variables in $\mathbf{Y}$ are removed in ascending order of their heights. It is obvious from the definition of height that 
        there are no directed paths from $Y_j$ to nodes with the same or greater height. However, the causal relationships for the other case are yet to be determined. The following proposition helps to complete the construction of $G^+$.
        
        \begin{proposition}
        \label{between}
        Suppose that Assumptions \ref{assumption-independent X}--\ref{assumption-valid IV} hold. For any $k\in\leaf(G^-)$ and $Y_j\in\mathbf{Y}\setminus\mathbf{Y}^-$:
        \begin{enumerate}[label=(\roman*)]
            \item if $X_\ell\nind Y_j$ for all $\ell\in\iv_{G^-}(k)$, then $(k,j)\in\mathcal{E}^+$;
            \item if $Y_k$ is an unmediated parent of $Y_j$, then $X_\ell\nind Y_j$ for all $\ell\in\iv_{G^-}(k)$.
        \end{enumerate}
        \end{proposition}
        
        Proposition \ref{between} allows us to derive the ancestral relationships between $\mathbf{Y}_{\leaf(G^-)}$ and $\mathbf{Y}\setminus\mathbf{Y}^-$ by
        \[
        \{(k,j):k\in\leaf(G^-),Y_j\in\mathbf{Y}\setminus\mathbf{Y}^-,X_\ell\nind Y_j\text{ for all }\ell\in\iv_{G^-}(k)\}\subseteq\mathcal{E}^+,
        \]
        which ensures that all edges from an unmediated parent to $Y_j$ are included. The remaining edges in $\mathcal{E}^+$ correspond to the directed paths containing mediators in $G$. Since these paths are formed by edges previously identified, adding the ancestral relationships inferred from these paths to $\mathcal{E}^+$ is sufficient to recover $\mathcal{E}^+$. Moreover, we can reconstruct $\mathcal{I}^+$ by
        \[
        \mathcal{I}^+=\{(\ell,j):\text{for some }k\in\an_G(j)\cup\{j\},X_\ell\nind Y_k\}.
        \]
        Subsequently, by Definition \ref{definition of candidate IV}, the candidate IV sets are identified by
        \[
        \ca_G(k)=\{\ell:(\ell,k)\in\mathcal{I}^+\text{ and }(\ell,j)\in\mathcal{I}^+,k\ne j\text{ only if }(k,j)\in\mathcal{E}^+\},\quad k=1,\dots,p.
        \]
        
        \subsection{Identification of causal effects}\label{sec:ident_effects}
        Building on the identifiability of $\mathcal{E}^+$ and $\{\ca_G(k)\}_{k=1}^p$ established in the preceding subsection, we now proceed to
        demonstrate the identification of the causal effects $\bm{\beta^*}=(\beta_{kj}^*)_{(k,j)\in\mathcal{E}^+}$ and subsequently identify $\mathcal{E}$.
        
        First, consider the simple case where $(k,j)\in\mathcal{E}^+$ and $\me_G(k,j)=\emptyset$. By Definition \ref{definition of IH} and the identifiability of $\ca_G(k)$, $\mathcal{Z}_\gamma(k)$ and hence $\mathbf{Z}_\gamma(\mathbf{X}_{\ca_G(k)})$ are also identifiable. From the discussion below \eqref{gmm-nonmediator}, we have 
        \begin{equation*}
        \label{GMM_validIV}
        \mathbb{E}\{\mathbf{Z}_\gamma(\mathbf{X}_{\ca_G(k)})(Y_j-\beta_{kj}^*Y_k)\}=\mathbf{0}.
        \end{equation*}
        Under Assumption \ref{assumption-L0-norm}, this equation has a unique solution, and thus $\beta^*_{kj}$ is identifiable. Based on the above analysis, we can identify all $\beta_{kj}^*$ with $l_G(k,j)=1$. 
        
        Next, we identify the remaining parameters recursively. Suppose we have identified all $\beta_{kj}^*$ with $l_G(k,j)\le l$ for some $l>0$. Then, for any $\beta_{kj}^*$ with $l_G(k,j)=l+1$, any mediator variable $Y_i\in\me_G(k,j)$ satisfies $l_G(i,j)\le l$, and thus all mediated effects $\beta_{ij}^*$ are identified. We can then substitute $Y_j-\sum_{i\in\me_G(k,j)}\beta_{ij}^*Y_i$ for $Y_j$ and identify $\beta_{kj}^*$ under Assumption \ref{assumption-L0-norm} from the equation $\mathbb{E}\{M_{kj}(\bm{\beta^*})\}=0$, where
        \[
        M_{kj}(\bm{\beta^*})=\mathbf{Z}_\gamma(\mathbf{X}_{\ca_G(k)})\Biggl(Y_j-\sum_{i\in\me_G(k,j)}\beta_{ij}^*Y_i-\beta_{kj}^*Y_k\Biggr).
        \]
        The above procedure implies that we can determine all $\beta_{kj}^*$ recursively in ascending order of $l_G(k,j)$. As a result, $\bm{\beta^*}$ can be identified as the unique solution to
        \begin{equation}\label{eq:moment}
        \mathbb{E}\{\bM(\bm{\beta^*})\}=\mathbf{0},
        \end{equation}
        where $\bM(\bm{\beta^*})$ is the concatenation of all $M_{kj}(\bm{\beta^*})$ for $(k,j)\in\mathcal{E}^+$. Finally, based on the identified value of $\bm{\beta^*}$, we can identify $\mathcal{E}$ by
        \[
        \mathcal{E}=\{(k,j):\beta_{kj}^*\ne 0,(k,j)\in\mathcal{E}^+\}.
        \]
        
        To summarize, our analysis in Sections \ref{sec:ident_graph} and \ref{sec:ident_effects} shows that the causal graph and causal effects in model \eqref{SEM} are identifiable under Assumptions \ref{assumption-independent X}--\ref{assumption-L0-norm}, thereby justifying Theorem \ref{thm:ident}. A complete proof can be found in Appendix \ref{sec:proofs}.
        
        \section{Methodology and theory}\label{sec:method}
        In this section, we introduce PLACID, a finite-sample method for estimating the causal graph and causal effects in model \eqref{SEM}. The method consists of an algorithm for estimating the ARG and candidate IV sets (Section \ref{sec:method_graph}) and a procedure for inferring the causal effects and directions (Section \ref{sec:method_effects}). Theoretical guarantees are provided in Section \ref{sec:theory}.
        
        \subsection{Estimation of the ARG and candidate IV sets}\label{sec:method_graph}
        To estimate the ARG, it is essential to establish the dependence between $\mathbf{X}$ and $\mathbf{Y}$ based on finite samples, which can be achieved using distance correlation \citep{szekely2007measuring}. The distance correlation (DC) measures the dependence between two random vectors using the distance between their characteristic functions. Unlike Pearson's correlation coefficient, the DC is zero only if the random vectors are independent. Moreover, in the bivariate normal case, it is strictly increasing with the absolute value of Pearson's correlation coefficient. The notion of DC has been effectively used in feature screening \citep{li2012feature} and causal discovery from time series \citep{runge2019detecting}. Here we employ the empirical DC to test the independence of any pair $X_i$ and $Y_j$. Consider an independent and identically distributed sample $(\mathbf{X}_{n\times q},\mathbf{Y}_{n\times p})$. The empirical DC between $X_i$ and $Y_j$ is defined by
        \begin{equation}
            \label{empirical DC}
            \mathcal{R}_n(X_i,Y_j)=\frac{\mathcal{V}_n(X_i,Y_j)}{\sqrt{\mathcal{V}_n(X_i,X_i)\mathcal{V}_n(Y_j,Y_j)}},
        \end{equation}
        where $\mathcal{V}_n(\cdot,\cdot)$ is the empirical distance covariance. Note from \citet[(2.18)]{szekely2007measuring} that $\mathcal{V}_n^2(X_i,Y_j)=S_1(X_i,Y_j)+S_2(X_i,Y_j)-2S_3(X_i,Y_j)$, where 
        \begin{align*}
            S_1(X_i,Y_j)&=\frac{1}{n^2}\sum_{r=1}^n\sum_{s=1}^n|X_{ri}-X_{si}|\,|Y_{rj}-Y_{sj}|,\\
            S_2(X_i,Y_j)&=\frac{1}{n^2}\sum_{r=1}^n\sum_{s=1}^n|X_{ri}-X_{si}|\frac{1}{n^2}\sum_{r=1}^n\sum_{s=1}^n|Y_{rj}-Y_{sj}|,\\
            S_3(X_i,Y_j)&=\frac{1}{n^3}\sum_{r=1}^n\sum_{s=1}^n\sum_{t=1}^n|X_{ri}-X_{ti}|\,|Y_{sj}-Y_{tj}|.
        \end{align*}
        The distance variances $\mathcal{V}_n(X_i,X_i)$ and $\mathcal{V}_n(Y_j,Y_j)$ are calculated similarly. For testing the null hypothesis $H_0\colon X_i\ind Y_j$, the test statistic is given by
        \begin{equation}
        \label{DC test}
            T_n(X_i,Y_j)=\frac{n\mathcal{V}_n^2(X_i,Y_j)}{S_2(X_i,Y_j)}.
        \end{equation}
        Then a test of asymptotic significance level $\alpha$ rejects $H_0$ when $\sqrt{T_n(X_i, Y_j)}>\Phi^{-1}(1-\alpha/2)$, where $\Phi(\cdot)$ is the standard normal cumulative distribution function \citep[Theorem 6]{szekely2007measuring}.
        
        Following our analysis in Section \ref{sec:ident_graph}, we propose Algorithm \ref{alg:peel} for estimating the ARG and candidate IV sets. According to Proposition \ref{leafnodes}, for each subgraph $G^-$, the algorithm first searches for a secondary variable that empirically depends on the fewest variables in $\mathbf{Y}^-$ to serve as an IV. It then identifies the variable in $\mathbf{Y}^-$ that is most strongly correlated with this secondary variable as a leaf node. Next, the empirical dependencies are used to recover the relationships between $\mathbf{Y}_{\leaf(G^-)}$ and $\mathbf{Y}\setminus\mathbf{Y}^-$ by Proposition \ref{between}. By updating the subgraph $G^-$ and repeating the above procedure, the algorithm ultimately reconstructs the ARG and all candidate IV sets.
        
\begin{algorithm}[t]
        \KwIn{Data $(\mathbf{X}_{n\times q},\mathbf{Y}_{n\times p})$, significance level $\alpha>0$}
        \KwOut{Estimates of $\mathcal{E}^+$, $\mathcal{I}^+$, and candidate IV sets}
        
        Compute $\bC=(C_{ij})_{q\times p}$ via \eqref{empirical DC}: $C_{ij}\gets\mathcal{R}_n(X_i,Y_j)$\;
        Compute $\bR=(R_{ij})_{q\times p}$ via \eqref{DC test}: $R_{ij}\gets\mathbf{1}\{\sqrt{T_n(X_i,Y_j)}>\Phi^{-1}(1-\alpha/2)\}$\;
        Initialize $\hat{\mathcal{E}}^+\gets\emptyset$, $\hat{\mathcal{I}}^+\gets\{(i,j):R_{ij}\ne 0\}$\;
        Initialize $\mathbf{Y}^-\gets\mathbf{Y}$, $\mathbf{X}^-\gets\mathbf{X}$, $\mathcal{E}^-\gets\hat{\mathcal{E}}^+$, $\mathcal{I}^-\gets\hat{\mathcal{I}}^+$, $\bR^-\gets \bR$\;
        \While{$\mathbf{Y}^-\ne\emptyset$}{
            Initialize $\leaf(G^-)\gets\emptyset$, $\iv_{G^-}(j)\gets\emptyset$ for all $Y_j\in\mathbf{Y}^-$\;
            \For{$\ell\in\argmin_{j:\|\bR^-_{j,\cdot}\|_0>0}\|\bR_{j,\cdot}^-\|_0$}{
                $k\gets\argmax_{j:Y_j\in\mathbf{Y}^-}C_{\ell j}$\;
                $\leaf(G^-)\gets\leaf(G^-)\cup\{k\}$\;
                $\iv_{G^-}(k)\gets\iv_{G^-}(k)\cup\{\ell\}$\;
            }
            $\hat{\mathcal{E}}^+\gets\hat{\mathcal{E}}^+\cup\{(k,j):k\in\leaf(G^-),Y_j\in\mathbf{Y}\setminus\mathbf{Y}^-,R_{\ell j}\ne 0\text{ for all }\ell\in\iv_{G^-}(k)\}$\;
            $\mathbf{Y}^-\gets\mathbf{Y}^-\setminus\mathbf{Y}_{\leaf(G^-)}$, $\mathbf{X}^-\gets\mathbf{X}^-\setminus\bigcup_{k\in\leaf(G^-)}\mathbf{X}_{\iv_{G^-}(k)}$\;
            Update $\bR^-$ by keeping the rows in $\mathbf{X}^-$ and columns in $\mathbf{Y}^-$\;
        }
        $\hat{\mathcal{E}}^+\gets\{(k,j):Y_k\to\dots\to Y_j\text{ in }\hat{\mathcal{E}}^+\}$\;
        $\hat{\mathcal{I}}^+\gets\{(\ell,j):(\ell,k)\in\hat{\mathcal{I}}^+\text{ and }(k,j)\in\hat{\mathcal{E}}^+\}$\;
        $\widehat{\ca}_G(k)\gets\{\ell:(\ell,k)\in{\hat{\mathcal{I}}^+}\text{ and }(\ell,j)\in{\hat{\mathcal{I}}^+},k\ne j\text{ only if }(k,j)\in{\hat{\mathcal{E}}^+}\}$ for $k=1,\dots,p$\;
        
        \Return{$\hat{\mathcal{E}}^+$, $\hat{\mathcal{I}}^+$, $\{\widehat{\ca}_G(k)\}_{k=1}^p$}
        \caption{DC-based estimation of the ARG and candidate IV sets}\label{alg:peel}
        \end{algorithm}
         
        \subsection{Estimation of causal effects}\label{sec:method_effects}
        The estimated ARG obtained from Algorithm \ref{alg:peel} provides an initial understanding of the causal structures among the variables in $\mathbf{Y}$. We now show how to estimate the causal effects and hence recovery the edges $\hat{\mathcal{E}}$ using the candidate IV sets.
        
        On the basis of the moment condition \eqref{eq:moment}, we propose to estimate $\bm{\beta^*}$ using the generalized method of moments (GMM) \citep{hansen1982large}. The first step is to construct estimates of the unknown functions $\mathbf{Z}_\gamma(\mathbf{X}_{\ca_G(k)})$ in $\bM(\bm{\beta^*})$. This can be done in different ways according to the types of candidate IVs. In the case where all variables in $\mathbf{X}_{\ca_G(k)}$ take values in $\{0,1\}$, we follow the idea of \citet{sun2023semiparametric} to construct $\mathbf{Z}_\gamma(\mathbf{X}_{\ca_G(k)})$. Let $\alpha(1),\dots,\alpha(t_k)$ be an enumeration of all subsets $\alpha\subseteq\ca_G(k)$ of cardinality $|\alpha|\ge|\ca_G (k)|-\gamma+1$. Clearly, for any such $\alpha$ and any subset $\alpha_k\subseteq\ca_G(k)$ with $|\alpha_k|\ge\gamma$, we have $\alpha\cap\alpha_k\ne\emptyset$. This implies that $\mathbb{E}\{\Pi_{s\in\alpha}(X_{s}-\alpha_{s})\mid\mathbf{X}_{\ca_G(k)\setminus \alpha_k}\}=0$, where $\mu_s=\mathbb{E}(X_s)$, and hence
        \[
        \mathbf{Z}_\gamma(\mathbf{X}_{\ca_G(k)})=\bigl(\Pi_{s\in\alpha(1)}(X_s-\mu_s),\dots,\Pi_{s\in\alpha(t_k)}(X_s-\mu_s)\bigr)^T.
        \]
        A specific example is given in Appendix \ref{sec:basis}. To estimate $\mathbf{Z}_\gamma(\mathbf{X}_{\ca_G(k)})$, one simply substitutes the empirical means $\hat{\mathbb{E}}_n(X_s)$ for $\mu_s$ in the above expression. Furthermore, if $\mathbf{X}_{\ca_G(k)}$ includes polytomous variables, we break them into dummy variables and compute $\mathbf{Z}_\gamma(\mathbf{X}_{\ca_G(k)})$ similarly. For the continuous case, we use the strategy discussed after Definition \ref{definition of IH} to approximate $\mathbf{Z}_\gamma(\mathbf{X}_{\ca_G(k)})$.
        
        After obtaining estimates of $\mathbf{Z}_\gamma(\mathbf{X}_{\ca_G(k)})$, we proceed with the GMM estimation of $\bm{\beta^*}$ and recovery of $\mathcal{E}$, as summarized in Algorithm \ref{alg:gmm}. Note that the weighting matrix ${\bOmega}$ in \eqref{GMM-algorithm} may affect the asymptotic variance of the GMM estimator. In practice, ${\bOmega}$ can be either specified as the identity matrix or computed from the data as suggested by \cite{hansen1982large}. In the next subsection, we derive the asymptotic normality of $\hat{\bbeta}$ (Theorem \ref{thm:asympt}). The result is used in Algorithm \ref{alg:gmm} to test whether the individual entries of $\bm{\beta^*}$ are zero, thereby allowing us to recover the edges in $\mathcal{E}$.  To adjust for multiple comparisons, we apply the Benjamini--Yekutieli method \citep{benjamini2001control} in Algorithm \ref{alg:gmm} to control the false discovery rate (FDR) at level $q^*$; see Theorem \ref{thm:fdr}.
        
\begin{algorithm}
        \KwIn{Data $(\mathbf{X}_{n\times q},\mathbf{Y}_{n\times p})$, ancestral edge set $\hat{\mathcal{E}}^+$, candidate IV sets $\{\widehat{\ca}_G(k)\}_{k=1}^p$, weighting matrix ${\bOmega}$, FDR level $q^*>0$}
        \KwOut{Estimates of $\bm{\beta^*}$ and $\mathcal{E}$}
        
        $N\gets|\hat{{\mathcal{E}}}^+|$\;
        For each $Y_k$ with descendants in $\hat{\mathcal{E}}^+$, obtain an estimate $\hat{\mathbf{Z}}_\gamma(\mathbf{X}_{\ca_G(k)})$ of $\mathbf{Z}_\gamma(\mathbf{X}_{\ca_G(k)})$\;
        For each $(k,j)\in\hat{\mathcal{E}}^+$, $\widehat{\me}_G(k,j)\gets\{i:(k,i)\in\hat{\mathcal{E}}^+,(i,j)\in\hat{\mathcal{E}}^+\}$\;
        For each $(k,j)\in\hat{\mathcal{E}}^+$ and $\bbeta=(\beta_{kj})\in\mathbb{R}^N$: $\hat{M}_{kj}(\bbeta)\gets\hatbasisk(Y_j-\sum_{i\in\widehat{\me}_G(k,j)}\beta_{ij}Y_i-\beta_{kj}Y_k)$\;
        Concatenate $\hat{M}_{kj}(\bbeta)$ into $\hat{\bM}(\bbeta)$ and solve the following problem:
        \begin{equation}\label{GMM-algorithm}
        \hat{\bbeta}\gets\argmin_{\bbeta}\hat{\mathbb{E}}_n\{\hat{\bM}(\bbeta)\}^T{\bOmega}\hat{\mathbb{E}}_n\{\hat{\bM}(\bbeta)\}
        \end{equation}
        
        Obtain the standard errors $\hat\sigma_{kj}$ of $\hat\beta_{kj}$ for all $(k,j)\in\mathcal{E}^+$ using Theorem \ref{thm:asympt}\;
        Calculate the $p$-values $P_{kj}\gets 2\{1-\Phi(|\hat\beta_{kj}|/\hat\sigma_{kj})\}$ for all $(k,j)\in\mathcal{E}^+$\;
        Order the $p$-values as $P_{(1)}\le\dots\le P_{(N)}$ with $P_{(i)}$ corresponding to $(k_i,j_i)\in\mathcal{E}^+$\;
        $\ell\gets\max\{i:P_{(i)}\le iq^*/(N\sum_{j=1}^Nj^{-1})\}$\;
        $\hat{\mathcal{E}}\gets\{(k_i,j_i)\}_{i=1}^\ell$\;
        
        \Return{$\hat\bbeta$, $\hat{\mathcal{E}}$}
        \caption{GMM estimation of $\bm{\beta^*}$ and $\mathcal{E}$}\label{alg:gmm}
        \end{algorithm}
        
        \subsection{Theoretical guarantees}\label{sec:theory}
        In this subsection, we provide theoretical guarantees for our PLACID method in terms of causal discovery and inference. We begin with the following result, showing that Algorithm \ref{alg:peel} consistently learns the ARG and candidate IV sets.
        
        \begin{theorem}[Consistency of ancestral structure recovery]
        \label{consistency of ARG}
        Suppose that Assumptions \ref{assumption-independent X}--\ref{assumption-valid IV} hold and $\alpha=O(n^{-2})$ in Algorithm \ref{alg:peel}. Then the estimated ARG $\hat{G}^+$ and candidate IV sets $\{\widehat{\ca}_G(k)\}_{k=1}^p$ from Algorithm \ref{alg:peel} satisfy
        \[
        \lim_{n\to\infty}\mathbb{P}(\hat{G}^+=G^+)=1
        \]
        and
        \[
        \lim_{n\to\infty}\mathbb{P}\{\widehat{\ca}_G(k)=\ca_G(k)\}=1,\quad k=1,\dots,p.
        \]
        \end{theorem}
        
        Assuming consistency of the estimates from Algorithm \ref{alg:peel}, we have the following results concerning the inference of causal effects and directions by Algorithm \ref{alg:gmm}.
        
        \begin{theorem}[Asymptotic normality of $\hat\bbeta$]
        \label{thm:asympt}
        Suppose that Assumptions \ref{assumption-independent X}--\ref{assumption-L0-norm} hold. Then the estimated causal effects $\hat\bbeta$ from Algorithm \ref{alg:gmm} satisfy
        \[
        \sqrt{n}(\hat\bbeta-\bm{\beta^*})\to_d N(\mathbf{0},\mathbf{V}),
        \]
        where the asymptotic variance
        \[
        \mathbf{V}=(\mathbf{0}_{q\times q},\bI_{|\mathcal{E}^+|})(\bG^T\bW_{\bOmega} \bG)^{-1}\bG^T\bW_{\bOmega} \bF\bW_{\bOmega} \bG(\bG^T\bW_{\bOmega} \bG)^{-1}(\mathbf{0}_{q\times q},\bI_{|\mathcal{E}^+|})^T,
        \]
        and the specific forms of $\bW_{\bOmega}$, $\bG$, and $\bF$ are given in Appendix \ref{sec:proofs}.
        \end{theorem}
        
        Note that Theorem \ref{thm:asympt} accounts for the uncertainty due to the unknown mean $\bm{\mu}=\mathbb{E}(\mathbf{X})$ in the construction of surrogate IVs. To this end, we augment the estimating equations for $\bbeta^*$ with those for $\bm{\mu}$. The asymptotic variance of $\hat\bbeta$ is then the corresponding submatrix of the usual sandwich estimator for the full asymptotic variance.
        
        For an estimated edge set $\hat{\mathcal{E}}$, let $\TP$, $\RE$, and $\FP$ denote the numbers of estimated edges with correct directions, those with reverse directions, and those not in the true DAG, respectively. Define the false discovery proportion of $\hat{\mathcal{E}}$ by
        \[
        \FDP(\hat{\mathcal{E}})=(\RE+\FP)/(\TP+\RE+\FP),
        \]
        and the false discovery rate of $\hat{\mathcal{E}}$ by $\FDR(\hat{\mathcal{E}})=\mathbb{E}\{\FDP(\hat{\mathcal{E}})\}$. The following result ensures that Algorithm \ref{alg:gmm} controls the FDR in edge recovery at the nominal level.
        
        \begin{theorem}[FDR control in edge recovery]
        \label{thm:fdr}
        Suppose that Assumptions \ref{assumption-independent X}--\ref{assumption-L0-norm} hold and $\mathcal{E}\ne\emptyset$. Then for any $q^*\in(0,1)$, the estimated edge set $\hat{\mathcal{E}}$ from Algorithm \ref{alg:gmm} satisfies
        \[
        \lim_{n\to\infty}\FDR(\hat{\mathcal{E}})\le q^*.
        \]
        \end{theorem}
        
        \section{Simulation studies}\label{sec:simul}
        This section examines the finite-sample performance of our PLACID method. For causal discovery, we compare our method with GrIVET \citep{chen2024discovery}, RFCI \citep{colombo2012learning}, and LRpS-GES \citep{frot2019robust}. Since the last two are unable to estimate the causal effects, we compare our method only with GrIVET for parameter estimation, where the effects of $\mathbf{X}$ on $\mathbf{Y}$ are specified in a linear form. We use the R package \texttt{grivet} for the implementation of GrIVET, \texttt{pcalg} for RFCI, and \texttt{lrpsadmm} and \texttt{pcalg} for LRpS-GES.
        
        We consider two types of DAGs with unobserved confounders: random graphs and hub graphs. Let $\bA\in\mathbb{R}^{p\times p}$ denote the adjacency matrix for the DAG. For random graphs, the upper off-diagonal entries of $\bA$ are independently sampled from $\text{Bernoulli}(1/(2p))$, while the other entries are set to 0. For hub graphs, the entries $A_{1j}$, $j=2,\dots,p$, are set to 1, with the remaining set to 0. Further, if $A_{kj}\ne0$, then $\beta^*_{kj}$ is sampled from the uniform distribution on $(-1.2,-0.8)\cup(0.8,1.2)$; otherwise, it is set to 0. We consider the structural equation model
        \[
        Y_j=\sum_{i=1}^p\beta_{ij}^*Y_i+g_j(\mathbf{X}_{\intr_G(j)})+\bphi_j^T\bU+e_j,\quad j=1,\dots,p,
        \]
        where $\mathbf{e}=(e_1,\dots,e_p)^T\sim N_p(\mathbf{0},\mathbf{\Lambda})$ and the unobserved confounders $\bU\sim N_r(\mathbf{0},\bI_r)$. Here $\mathbf{\Lambda}=\diag(\sigma_1^2,\dots,\sigma_p^2)$ with $\sigma_i$ sampled uniformly from $(0.3,0.4)$. The coefficients $\bphi_j=(\phi_{1j},\dots,\phi_{rj})^T$ are set as follows: $\phi_{11}$ and $\phi_{kj}$, $j=2k-1,2k$, $k=1,\dots,r$, are sampled uniformly from $(-0.4,-0.3)\cup(0.3,0.4)$, while the other entries are set to 0. We set $q=2p+\lfloor p/2 \rfloor$ and $\intr_G(j)=\{j,p+j,2p+\lfloor j/2\rfloor\}$. Hence, each $X_\ell$, $\ell=2p,\dots,2p+\lfloor(p-1)/2\rfloor$, intervenes on two primary variables, while any other $X_\ell$ intervenes on a single $Y_j$.
        
        We consider two types of secondary variables $\mathbf{X}$: (i) the continuous case where $X_i\sim N(0,1)$ independently, and (ii) the discrete case where $X_i\sim\text{Bernoulli}(0.5)$ independently. Depending on the types of $\mathbf{X}$, we specify $g_j(\mathbf{X}_{\intr_G(j)})$ as follows: for the continuous case,
        \[
        g_j(\mathbf{X}_{\intr_G(j)})=w_j\sum_{\ell\in\intr_G(j)}\{X_\ell^2+\mathbf{1}(X_\ell >0)\}+\frac{w_j}{2}\sum_{k,\ell\in\intr_G(j),\,k\ne\ell}X_kX_\ell,
        \]
        where $w_j$ are sampled uniformly from $(-3.2,-2.8)\cup(2.8,3.2)$; for the discrete case,
        \[
        g_j(\mathbf{X}_{\intr_G(j)})=\sum_{k,\ell\in\intr_G(j),\,k\ne\ell}X_kX_\ell.
        \]
        To illustrate the performance of our method for DAGs of different sizes, we fix the sample size at $n=1000$ while varying the dimension $p$ and adjusting $q$ and $r$ accordingly as $(p,q,r)=(10,25,5)$ and $(20,50,10)$. In Algorithm \ref{alg:gmm}, we set the weighting matrix ${\bOmega}=\bI$ and the FDR level $q^*=0.05$. For the continuous case, we use tensor products of polynomial bases to approximate $\mathbf{Z}_\gamma(\mathbf{X}_{\ca_G(k)})$.
        
        For causal discovery, RFCI outputs a partial ancestral graph and LRpS-GES outputs a completed partially DAG, both of which may include undirected edges. We evaluate these two methods favorably by assuming that the correct directions were obtained for undirected edges, as in \cite{li2024nonlinear}. Four performance metrics for causal discovery are used: false discovery proportion (FDP), true positive rate (TPR), structural Hamming distance (SHD), and Jaccard index (JI). Let $\TP$, $\RE$, and $\FP$ be defined as in Section \ref{sec:theory}, and $\FN$ the number of missing edges from the true DAG. Then $\FDP=(\RE+\FP)/(\TP+\RE+\FP)$, $\TPR=\TP/(\TP+\FN)$, $\SHD=\FP+\FN+\RE$, and $\JI=\TP/(\TP+\SHD)$. The simulation results for the continuous and discrete cases are summarized in Tables \ref{continuous_structure} and \ref{discrete_structure}, respectively.
        
        \begin{table}
        \centering
        \caption{Comparison of causal discovery performance for four competing methods with continuous secondary variables. Means and standard deviations (in parentheses) of performance metrics based on 100 replications are shown.}\label{continuous_structure}
        \def~{\phantom{0}}
        \begin{tabular*}{\textwidth}{@{}l*{2}{@{\extracolsep{\fill}}l}*{4}{@{\extracolsep{\fill}}c}@{}}
        \toprule
        Graph  & $p$ & Method   & FDP        & TPR        & SHD          & JI\\
        \midrule
        Random & 10  & PLACID   & 0.02(0.08) & 0.92(0.21) &  0.30(0.67)  & 0.90(0.22)\\
               &     & GrIVET   & 0.57(0.36) & 0.49(0.39) &  4.06(2.90)  & 0.27(0.27)\\
               &     & RFCI     & 0.07(0.24) & 0.48(0.39) &  1.31(1.29)  & 0.47(0.39)\\
               &     & LRpS-GES & 0.70(0.12) & 0.97(0.13) &  5.04(0.88)  & 0.30(0.12)\\
               & 20  & PLACID   & 0.02(0.07) & 0.91(0.15) &  0.58(0.98)  & 0.90(0.16)\\
               &     & GrIVET   & 0.83(0.15) & 0.43(0.26) & 16.38(9.70)~ & 0.14(0.11)\\
               &     & RFCI     & 0.03(0.13) & 0.59(0.28) &  2.07(1.60)  & 0.59(0.28)\\
               &     & LRpS-GES & 0.70(0.08) & 0.99(0.04) & 10.55(1.40)~ & 0.30(0.08)\\
        Hub    & 10  & PLACID   & 0.00(0.00) & 1.00(0.00) &  0.00(0.00)  & 1.00(0.00)\\
               &     & GrIVET   & 0.40(0.36) & 0.48(0.38) &  7.55(5.51)  & 0.41(0.36)\\
               &     & RFCI     & 0.01(0.04) & 0.58(0.21) &  3.78(1.89)  & 0.58(0.21)\\
               &     & LRpS-GES & 0.39(0.05) & 0.83(0.10) &  6.30(1.40)  & 0.55(0.09)\\
               & 20  & PLACID   & 0.00(0.00) & 1.00(0.00) &  0.00(0.00)  & 1.00(0.00)\\
               &     & GrIVET   & 0.65(0.31) & 0.41(0.36) & 25.82(13.41) & 0.28(0.28)\\
               &     & RFCI     & 0.07(0.09) & 0.45(0.15) & 11.04(3.00)~ & 0.44(0.15)\\
               &     & LRpS-GES & 0.43(0.04) & 0.80(0.08) & 15.38(2.14)~ & 0.50(0.06)\\
        \bottomrule
        \end{tabular*}
        \end{table}
        
        \begin{table}
        \centering
        \caption{Comparison of causal discovery performance for four competing methods with discrete secondary variables. Means and standard deviations (in parentheses) of performance metrics based on 100 replications are shown.}\label{discrete_structure}
        \def~{\phantom{0}}
        \begin{tabular*}{\textwidth}{@{}l*{2}{@{\extracolsep{\fill}}l}*{4}{@{\extracolsep{\fill}}c}@{}}
        \toprule
        Graph  & $p$ & Method   & FDP        & TPR        & SHD         & JI\\
        \midrule
        Random & 10  & PLACID   & 0.01(0.04) & 0.92(0.07) & ~0.07(0.30) & 0.91(0.08)\\
               &     & GrIVET   & 0.14(0.24) & 0.63(0.35) & ~1.08(1.15) & 0.57(0.34)\\
               &     & RFCI     & 0.00(0.00) & 0.82(0.26) & ~0.25(0.57) & 0.82(0.26)\\
               &     & LRpS-GES & 0.62(0.16) & 0.92(0.04) & ~3.39(1.14) & 0.38(0.16)\\
               & 20  & PLACID   & 0.03(0.07) & 0.98(0.05) & ~0.33(0.73) & 0.95(0.09)\\
               &     & GrIVET   & 0.28(0.24) & 0.65(0.25) & ~3.26(2.42) & 0.50(0.23)\\
               &     & RFCI     & 0.00(0.02) & 0.94(0.11) & ~0.34(0.80) & 0.94(0.11)\\
               &     & LRpS-GES & 0.64(0.12) & 0.97(0.06) & ~7.91(1.86) & 0.36(0.12)\\
        Hub    & 10  & PLACID   & 0.00(0.00) & 0.96(0.07) & ~0.39(0.63) & 0.96(0.07)\\
               &     & GrIVET   & 0.00(0.03) & 0.50(0.17) & ~4.55(1.56) & 0.50(0.17)\\
               &     & RFCI     & 0.00(0.00) & 0.44(0.21) & ~5.06(1.86) & 0.44(0.21)\\
               &     & LRpS-GES & 0.44(0.05) & 0.78(0.05) & ~7.52(1.11) & 0.49(0.05)\\
               & 20  & PLACID   & 0.00(0.00) & 0.97(0.04) & ~0.54(0.82) & 0.97(0.04)\\
               &     & GrIVET   & 0.01(0.10) & 0.49(0.13) & ~9.92(3.59) & 0.49(0.13)\\
               &     & RFCI     & 0.00(0.00) & 0.38(0.13) & 11.87(2.48) & 0.38(0.13)\\
               &     & LRpS-GES & 0.50(0.05) & 0.80(0.09) & 18.96(3.07) & 0.45(0.06)\\
        \bottomrule
        \end{tabular*}
        \end{table}
        
        Tables \ref{continuous_structure} and \ref{discrete_structure} indicate that PLACID performs best in terms of causal discovery across all scenarios. In particular, it effectively controls the FDP below the nominal level $q^*=0.05$, while maintaining a TPR higher than 0.9 for powerful edge detection. As expected, GrIVET struggles with nonlinear relationships between the primary and secondary variables, whereas RFCI and LRpS-GES are less effective in handling large effects of unobserved confounders. In the continuous case, PLACID shows remarkable accuracy for hub graphs, likely because the dependence between $\mathbf{X}$ and $\mathbf{Y}$ is well captured by the empirical DC in these settings. In the discrete case, the performance of RFCI is comparable to that of PLACID for random graphs; however, these metrics are calculated by assuming correct directions for undirected edges in RFCI, giving it an unfair advantage. Moreover, RFCI tends to be less powerful for hub graphs, while PLACID consistently exhibits superior and stable performance across different graph structures and sizes.
        
        For parameter estimation, we compare our method with GrIVET in terms of entrywise $L_\infty$, $L_1$, and $L_2$ losses, as reported in Table \ref{parameter_nonlinear}. These results demonstrate the superior performance of PLACID over GrIVET in parameter estimation across various settings. It is also interesting to note that the performance of PLACID in causal discovery and parameter estimation tend to have the same trend, as can be seen from a comparison between Tables \ref{continuous_structure} and \ref{discrete_structure} and Table \ref{parameter_nonlinear}. For example, in the continuous case with hub graphs, PLACID achieves minimal error and high accuracy in both causal discovery and parameter estimation. This is reasonable since a better estimate of the ARG enables more accurate parameter estimation, which in turn leads to a more precise recovery of causal structures.
        
        \begin{table}
        \centering
        \caption{Comparison of parameter estimation performance for two competing methods with continuous and discrete secondary variables. Means and standard deviations (in parentheses) of estimation losses based on 100 replications are shown.}\label{parameter_nonlinear}
        \def~{\phantom{0}}
        \begin{tabular*}{\textwidth}{@{}l*{3}{@{\extracolsep{\fill}}l}*{3}{@{\extracolsep{\fill}}c}@{}}
        \toprule
        Setting    & Graph  & $p$ & Method & $L_\infty$ & $L_1$       & $L_2$\\
        \midrule
        Continuous & Random & 10  & PLACID & 0.23(0.38) & ~0.35(0.67) & 0.26(0.46)\\
                   &        &     & GrIVET & 0.78(0.39) & ~1.91(1.45) & 1.05(0.62)\\
                   &        & 20  & PLACID & 0.37(0.44) & ~0.68(0.99) & 0.45(0.57)\\
                   &        &     & GrIVET & 1.07(0.17) & ~5.03(1.87) & 1.84(0.49)\\
                   & Hub    & 10  & PLACID & 0.11(0.03) & ~0.39(0.06) & 0.17(0.03)\\
                   &        &     & GrIVET & 0.99(0.29) & ~7.10(4.03) & 2.27(1.00)\\
                   &        & 20  & PLACID & 0.09(0.02) & ~0.63(0.11) & 0.17(0.03)\\
                   &        &     & GrIVET & 1.14(0.13) & 18.81(9.02) & 3.82(1.27)\\
        Discrete   & Random & 10  & PLACID & 0.12(0.21) & ~0.20(0.45) & 0.14(0.28)\\
                   &        &     & GrIVET & 0.76(0.38) & ~1.47(1.08) & 0.95(0.55)\\
                   &        & 20  & PLACID & 0.21(0.36) & ~0.40(0.71) & 0.26(0.45)\\
                   &        &     & GrIVET & 0.98(0.20) & ~3.26(1.60) & 1.56(0.50)\\
                   & Hub    & 10  & PLACID & 0.38(0.35) & ~0.93(0.57) & 0.48(0.37)\\
                   &        &     & GrIVET & 1.16(0.03) & ~9.00(0.34) & 3.02(0.11)\\
                   &        & 20  & PLACID & 0.36(0.22) & ~2.31(1.35) & 0.95(0.60)\\
                   &        &     & GrIVET & 1.18(0.02) & 19.99(0.53) & 4.50(0.12)\\
        \bottomrule
        \end{tabular*}
        \end{table}
        
        \section{Application to ADNI data}\label{sec:appl}
        Inferring gene regulatory networks is crucial for understanding the pathophysiology of complex diseases and developing effective therapeutics \citep{barabasi2011network}. In this section, we apply our method to the Alzheimer's Disease Neuroimaging Initiative (ADNI) dataset (https://adni.loni.usc.edu) for estimating gene regulatory networks. For our analysis, we use the preprocessed data from \cite{chen2024discovery}, with gene expression levels normalized and adjusted for baseline covariates. By selecting genes with at least one strongly associated single nucleotide polymorphism (SNP) and two strongest SNPs for each gene, the dataset includes $p=21$ genes as primary variables and $q=42$ SNPs as secondary variables. The individuals in the study were divided into two groups: 462 cases with Alzheimer's disease or mild cognitive impairment (AD-MCI), and 247 controls that are cognitively normal (CN). Partial residual plots in Appendix \ref{sec:partial} suggest the presence of nonlinear relationships between some primary and secondary variables, and hence model \eqref{SEM} is appropriate. We then apply PLACID to learn the DAGs among the genes for both groups.
        
        \begin{figure}
        \centering
        \subfloat[AD-MCI]{\label{AD}\includegraphics[width=0.5\textwidth]{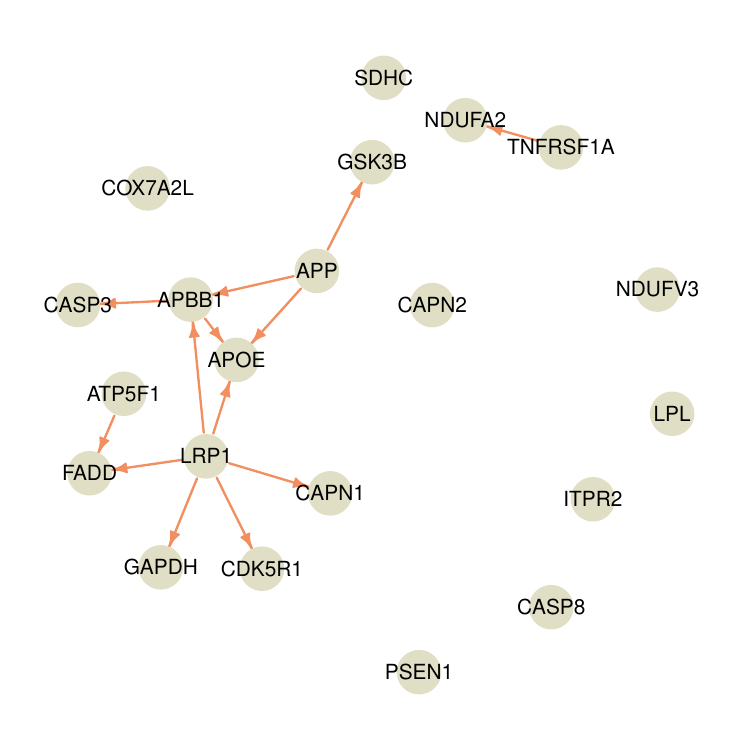}}
        \subfloat[CN]{\label{CN}\includegraphics[width=0.5\textwidth]{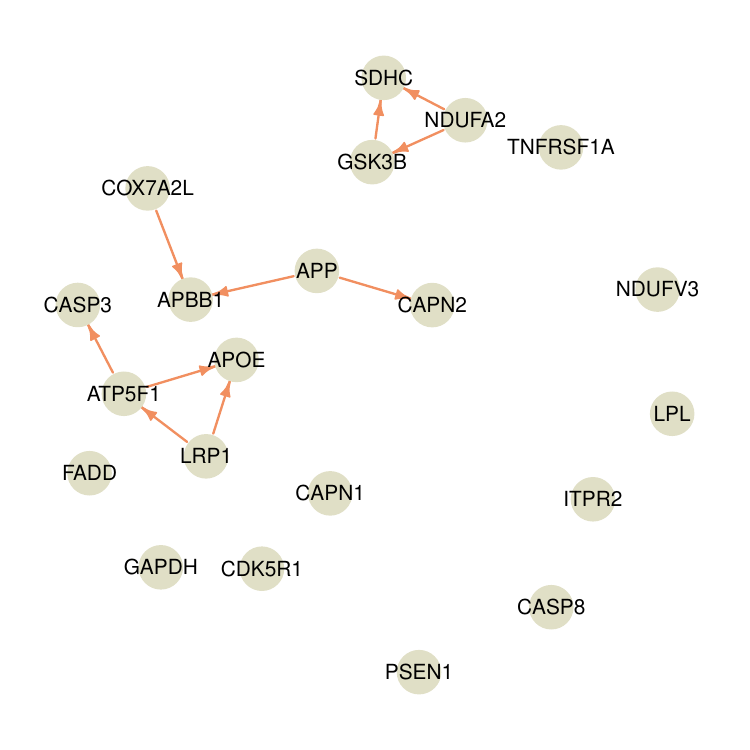}}
        \caption{Estimated gene regulatory networks for the (a) AD-MCI and (b) CN groups.}\label{fig:real}
        \end{figure}
        
        The estimated DAGs for the AD-MCI and CN groups are displayed in Figure \ref{fig:real}, which reveal both common and distinctive features of gene regulatory interactions in the two groups. Compared with the CN group, the AD-MCI group has substantially more edges originating from LRP1, suggesting a critical role of LRP1 in the pathogenesis of Alzheimer's disease. Indeed, it has been known that LRP1 is a major regulator of amyloid-$\beta$ and tau, the two hallmark proteins in Alzheimer's disease \citep{bloom2014amyloid}, and contributes to their accumulation and spread in the brain \citep{storck2016endothelial,rauch2020lrp1}. Among the outgoing edges of LRP1, the link to APOE is shared by both groups, which is consistent with the previous finding that LRP1 regulates brain APOE and cholesterol metabolism \citep{liu2007amyloid}. In fact, APOE has long been established as the strongest genetic risk factor for late-onset Alzheimer's disease, and has multifaceted effects on many neurobiological processes underlying Alzheimer's disease \citep{serrano2021apoe}. We further note that APP is connected to different downstream genes between the two groups. In particular, the AD-MCI group includes paths from APP to APOE, GSK3B, and APBB1, whereas the first two paths are absent in the CN group. APP is the precursor to amyloid-$\beta$, whose abnormal processing has been found central to the development of Alzheimer's disease \citep{o2011amyloid}. Interestingly, the paths from APP to APOE support the possibility that APOE and cholesterol levels are modulated, directly or indirectly, by APP \citep{liu2007amyloid}. In addition, GSK3B is involved in pathways that drive tau phosphorylation and synaptic dysfunction in Alzheimer's disease \citep{choi2014three}. Together, our results seem to have uncovered more edges, even after multiple testing adjustment, than those reported by \citet{chen2024discovery} using GrIVET, highlighting the potential of PLACID for practical causal discovery and inference. 
        
        \section{Discussion}\label{sec:disc}
        To conclude, we have proposed a novel method for the identification and inference of DAGs under unobserved confounding using invalid IVs. Still, our method may suffer from certain limitations and can be extended and improved in several directions. First, Assumption \ref{assumption-independent X} may be relaxed to allow dependence among the secondary variables. To block the paths through correlated secondary variables, one can apply the notion of conditional distance correlation \citep{wang2015conditional} to test the independence of $X_i$ and $Y_j$ conditional on the other secondary variables. For parameter estimation, one may follow the strategy of \cite{sun2023semiparametric} and correct the moment conditions with weights that account for possible dependence. Second, it would be valuable to extend our setting to nonlinear causal models with unobserved confounders, similar to those considered by \citet{agrawal2023decamfounder} and \citet{li2024nonlinear}. Finally, in high-dimensional settings, it is of interest to investigate whether the estimation of the ARG remains consistent and how this uncertainty affects subsequent parameter estimation. We leave these topics for future research.
        
        
        \appendix\label{Appendix}
        
        \section{An example of the surrogate IVs}\label{sec:basis}
        
        Since the estimation of  $\mathbf{Z}_\gamma(\mathbf{X}_{\ca_G(k)})$ plays an important role in our estimates for $\bm{\beta}^*$, we further clarify its concrete instantiation through Example \ref{example IH}. 
        
        \begin{example}
        \label{example IH}
        Focus on $k=2$ in Figure \ref{example 1-figure} where $\gamma=1$ according to Assumption \ref{assumption-valid IV}. The candidate IV set for $Y_2$ is $\{2,4\}$ and suppose that $X_i\in \{0,1\},\ i=2,4$. Then we have
        \[
        \mathcal{H}(\mathbf{X}_{\{2,4\}})=\operatorname{span}(\{X_2-\mu_2,X_4-\mu_4,(X_2-\mu_2)(X_4-\mu_4)\}),
        \]
        where $\mu_i=\mathbb{E}(X_i),\ i=2,4$. The valid IV set for $Y_2$ which satisfies $\iv_G(2)\subseteq \ca_G(2)$ and $|\iv_G(2)|\ge 1$ can be $\{2\},\ \{4\}$ or $\ \{2,4\}$, corresponding to the potential values for $\alpha_2$ in Definition \ref{definition of IH}. According to Assumption \ref{assumption-independent X}, when $\alpha_2=\{2\}$, the corresponding $\mathcal{D}(\{2\})$ is $\operatorname{span}(\{X_2-\mu_2,(X_2-\mu_2)(X_4-\mu_4)\})$; when $\alpha_2=\{4\}$,  $\mathcal{D}(\{4\})=\operatorname{span}(\{X_4-\mu_4,(X_2-\mu_2)(X_4-\mu_4)\})$; and when $\alpha_2=\{2,4\}$, $\mathcal{D}(\{2,4\})=\operatorname{span}(\{X_2-\mu_2,X_4-\mu_4,(X_2-\mu_2)(X_4-\mu_4)\})$. The intersection of all possible $\mathcal{D}(\alpha_2)$, defined as $\mathcal{Z}_1(2)$, is $\operatorname{span}(\{(X_2-\mu_2)(X_4-\mu_4)\})$. Therefore, the vector $\mathbf{Z}_{1}(\mathbf{X}_{\ca_G(2)})$ consisting of the basis functions of $\mathcal{Z}_1(2)$ is $((X_2-\mu_2)(X_4-\mu_4))$.
        \end{example}
        
        \section{Technical proofs}\label{sec:proofs}
        
        We begin by establishing the proof of Proposition \ref{leafnodes} and Proposition \ref{between}, which will be used in the proof of Theorem \ref{thm:ident}.
        
        \subsection{Proof of Proposition \ref{leafnodes}}
        
        \begin{proof}
            On the one hand, if $Y_k$ is a leaf node of ${G}$, then let $X_\ell$ be a valid IV of $Y_k$ since $\iv(k)\ne \emptyset$ due to Assumption \ref{assumption-valid IV}. We then have $X_\ell\ind Y_{k'}$ for any $k'\ne k$ because $Y_k$ has no descendant and $X_\ell \ind X_{i}$ for any $i\ne \ell$ according to Assumption \ref{assumption-independent X}. On the other hand, if $Y_k$ is not a leaf node of $G$, then there exists a primary variable $Y_j$ such that $Y_k$ is an unmediated parent of $Y_j$. There are two cases leading to $X_\ell\not\ind Y_k$, i.e., $(\ell,k)\in \mathcal{I}$ and $(\ell,k)\notin\mathcal{I}$. If $(\ell,k)\in\mathcal{I}$, then we have $X_\ell\not\ind Y_j$ according to Assumption \ref{assumption-unmediated parent}. If $(\ell,k)\notin\mathcal{I}$, then there exists an ancestor $i\in\an_G(k)$ such that $(\ell,i)\in\mathcal{I}$, and thus $X_\ell\not\ind Y_i$.  Therefore, for any non-leaf of $G$, we can not find an $X_\ell$ satisfying the conditions in Proposition \ref{leafnodes}. The analysis above shows the identification of leaf nodes in $G$. Moreover, under Assumptions \ref{assumption-independent X}--\ref{assumption-valid IV}, for any leaf node $Y_k$ of $G$, it follows that $X_\ell\nind Y_k$ and $X_\ell\ind Y_{k'}$ for all $k'\ne k$ if and only if $\ell\in \iv_G(k)$. 
        \end{proof}
        
        \subsection{Proof of Proposition \ref{between}}
        
        \begin{proof}
        It is obvious that $\iv_G(k)\subseteq \iv_{G^-}(k)$ for any $Y_k\in \mathbf{Y}^-$ according to Definition \ref{definition of valid IV} for valid IVs. According to Assumption \ref{assumption-independent X} and Definition \ref{definition of valid IV}, if $X_\ell\! \ind Y_k$ for all $\ell\in \iv_G(k)$, there must be a directed path from $Y_k$ to $Y_j$ in $G$. Thereby, we prove the first conclusion of Proposition \ref{between}. Moreover, for any $\ell\in \iv_{{G}^-}(k)$, we have $(\ell,k)\in \mathcal{I}^-\subseteq\mathcal{I}$ based on the definition of valid IV in Definition \ref{definition of valid IV}. Therefore, if $Y_k$ is an unmediated parent of $Y_j$, we have $X_\ell \nind Y_j$ according to Assumption \ref{assumption-unmediated parent}, which leads to the second conclusion.
        \end{proof}

        \subsection{Proof of Theorem \ref{thm:ident}}
        
        \begin{proof}
        Our proof consists of two parts. First, we prove the identifiability of $G^+$ and $\ca_{{G}(k)},k=1,\dots,p$. Under Assumptions \ref{assumption-independent X}--\ref{assumption-valid IV}, we can identify the leaves of $G$, denoted $\mathbf{Y}_{L(1)}$, according to Proposition \ref{leafnodes}. After removing $\mathbf{Y}_{L(1)}$ and its IVs to get a subgraph $G^-$, Proposition \ref{leafnodes} remains valid since Assumptions \ref{assumption-independent X}--\ref{assumption-valid IV} continue to hold for $G^-$. Therefore, we can identify the leaves of $G^-$, denoted as $\mathbf{Y}_{L(2)}$. Repeating this procedure, we can reorganize the primary variables into $H$ parts by their height, namely $\mathbf{Y}_{L(1)},\dots,\mathbf{Y}_{L(H)}$. For any $h^{\prime}\ge h$, it is evident that none of $\mathbf{Y}_{L({h^{\prime}})}$ is a descendant of $\mathbf{Y}_{L({h})}$. Therefore, it suffices to identify which variables in $\cup_{s<h}\mathbf{Y}_{L(s)}$ are descendants of $\mathbf{Y}_{L(h)}$. Let $G^-(h)$ be the subgraph whose leaves are $\mathbf{Y}_{L(h)}$. Then we can identify parts of $\mathcal{E}^+$ according to Proposition \ref{between} as: 
        \[
        \mathcal{E}^+_0=\bigcup_{h=2}^H\bigl\{(k,j):k\in L(h),\ j\in \cup_{s<h}L(s), \text{ and }\ X_\ell \not\ind Y_j \text{ for all } \ell\in \iv_{G^-(h)}(k)\bigr\}.
        \]
        Moreover, Proposition \ref{between} implies that $\mathcal{E}^+_0$ includes all edges originating from unmediated parents, and no additional edges are incorrectly included. Since all directed paths in $G$ can be connected by edges originating from their corresponding unmediated parents, we can reconstruct $\mathcal{E}^+$ as:
        \[
        \mathcal{E}^+=\{(k,j): \text{there is a directed path from } Y_k \text{ to } Y_j \text{ in } \mathcal{E}^+_0\}.
        \]
        For $\mathcal{I}^+$, we can construct it based on the identified $\mathcal{E}^+$ by using the relationships between $\mathbf{X}$ and $\mathbf{Y}$ as follows:
            \[
            {\mathcal{I}}^+=\bigl\{(\ell,j):\ \text{for some } i \in \{j\}\cup\{i:(i,j)\in \mathcal{E}^+\},\ \text{such that}\ X_\ell\not\ind Y_{i}\bigr\}.
            \]
        Therefore, the ARG $G^+$ is identifiable. Then, according to Definition \ref{definition of candidate IV}, the candidate IV sets can be identified as follows:
        \[
        \ca_G(k)=\{\ell:(\ell,k)\in {\mathcal{I}^+}\text{ and } (\ell,j)\in{\mathcal{I}^+},\ j\neq k\text{ only if }(k,j)\in{\mathcal{E}^+}\},\ 1\le k\le p.
        \]
        
        Next, we show that the causal effects $\bB^*$ are also identifiable based on the identified ARG and candidate IV sets. Since $\beta^*_{ij}=0$ if $(i,j)\notin \mathcal{E}^+$, it suffices to prove the identification of $\bm{\beta^*}=(\beta_{kj}^*)_{(k,j)\in\mathcal{E}^+}$. We first establish the moment conditions for $\bm{\beta^*}$. For any $(k,j)\in\mathcal{E}^+$, it is easy to see that $\mathbf{Y}_{\nm_G(k,j)}\ind \mathbf{X}_{\ca_G(k)}$ and $\mathbf{X}_{\intr_G(j)}\ind \mathbf{X}_{\ca_G(k)}\mid \mathbf{X}_{\ca_G(k)\setminus \iv_G(k)}$. Therefore, for any $d(\mathbf{X}_{\ca_G(k)})\in \mathcal{D}(\iv_G(k))$, we have
        \[
        \begin{aligned}
                &\mathbb{E}\{d(\mathbf{X}_{\ca_G(k)})(Y_j-\beta^*_{kj}Y_k)\}\\
                &\quad=\mathbb{E}\Biggl\{d(\mathbf{X}_{\ca_G(k)})\Biggl(\sum_{i\in \nm_G(k,j)}\beta^*_{ij}Y_i+g_j(\mathbf{X}_{\intr_G(j)})+\ve_j\Biggr)\Biggr\}\\
                &\quad=\mathbb{E}\{d(\mathbf{X}_{\ca_G(k)})g_j(\mathbf{X}_{\intr_G(j)})\}\\
                &\quad=\mathbb{E}\bigl[\mathbb{E}\{d(\mathbf{X}_{\ca_G(k)})\mid\mathbf{X}_{\ca_G(k)\setminus \iv_G(k)}\} \mathbb{E}\{g_j(\mathbf{X}_{\intr_G(j)})\mid \mathbf{X}_{\ca_G(k)\setminus\iv_G(k)}\}\bigr]\\
                &\quad=0.
        \end{aligned}
        \]
        Since $\mathcal{Z}_\gamma(k)\subseteq \mathcal{D}(\iv_G(k))$, it follows that
        \begin{equation}
            \label{eq:gmm-general}  
        \mathbb{E}\Biggl\{\mathbf{Z}_\gamma(\mathbf{X}_{\ca_G(k)}) \Biggl(Y_j-\beta_{kj}^*Y_k-{\sum_{i\in \me_G(k,j)}}\beta^*_{ij}Y_i\Biggr)\Biggr\}=\mathbf{0}.
        \end{equation}
        We next prove the identifiability of $\bm{\beta^*}$ using mathematical induction based on the value of $l_G(k,j)$, which is also identifiable due to the identified ARG. 
        
        \begin{enumerate}[label={(\roman*)}]
            \item All $\beta^*_{kj}$'s with $ (k,j)\in\mathcal{E}^+\text{ and }l_G(k,j)=1$ are identifiable. According to \eqref{eq:gmm-general}, for all $(k,j)\in \{(k,j):\ (k,j)\in\mathcal{E}^+\text{ and }l_G(k,j)=1\}$, we have
        \[
                \mathbb{E}\big\{\mathbf{Z}_\gamma(\mathbf{X}_{\ca_G(k)})(Y_j-\beta^*_{kj}Y_k)\big\}=\mathbf{0}.
        \] 
            Under Assumption \ref{assumption-L0-norm}, the above equation has a unique solution $\beta^*_{kj}$. Thus, the parameters in the set $\{\beta^*_{kj}:(k,j)\in\mathcal{E}^+\text{ and }l_G(k,j)=1\}$ are identifiable.
        
            \item If $\max_{j=1,\dots,p}h_G(j)=1$, it is evident that $\bm{\beta^*}$ is identifiable. Otherwise, assume that all $\beta^*_{kj}$'s with $l_G(k,j)\le l$ are identifiable for some $l$ smaller than $\max_{j=1,\dots,p}h_G(j)$.
            
            \item Next we prove that all $\beta^*_{kj}$'s with $l_G(k,j)= l+1$ are identifiable. For any $\beta^*_{kj}$ with $l_G(k,j)= l+1$, each mediator $i\in \me_G(k,j)$ satisfies $\ l_G(i,j)\le l$, and thus all the mediated effects of $Y_k$ on $Y_j$ are identifiable. In light of the identified $\{\beta^*_{ij}\}_{i\in \me_G(k,j)}$, we can identify $\beta_{kj}^*$ as the unique solution to \eqref{eq:gmm-general} under Assumption \ref{assumption-L0-norm}. Therefore, all $\beta^*_{kj}$'s with $l_G(k,j)= l+1$ are identifiable.
            \end{enumerate}
            
            By mathematical induction, we conclude that $\bm{\beta^*}$ is identifiable. It follows that the causal effects $\bB^*$ among the primary variables are identifiable. In other words, if we define $\mathbf{M}(\bm{\beta^*})$ in the same way as in the main text, we have $\mathbb{E}\{\bM(\bm{\beta^*})\}=\mathbf{0}$ regardless of the exact valid IV sets. Assumption \ref{assumption-L0-norm} can be seen as a sufficient condition to make $\mathbb{E}\{\partial \bM(\bbeta)/\partial \bbeta\}$ full rank for any $\bbeta\in\mathbb{R}^N$. Therefore, we can identify $\bm{\beta^*}$ as the unique solution to $\mathbb{E}\{\bM(\bbeta)\}=\mathbf{0}$. Moreover, since $\mathcal{E}\subseteq\mathcal{E}^+$ and $(k,j)\in \mathcal{E}$ if and only if $\beta^*_{kj}\ne 0$, we can identify the edges among $\mathbf{Y}$ as $\mathcal{E}=\{(k,j):\beta^*_{kj}\ne 0,\ (k,j)\in \mathcal{E}^+\}$.
            \end{proof}
        
        \subsection{Proof of Theorem \ref{consistency of ARG}}
        
        \textbf{Distance correlation}
        
        We first introduce the distance correlation (DC) proposed by \cite{szekely2007measuring} in a brief. Let $f_{\zeta}(\cdot)$ denote the characteristic function for a random variable $\zeta$. 
        
        \begin{definition}[Distance correlation]
             The distance correlation between random vectors ${\bm{\eta}}$ and ${\bm{\xi}}$ with finite first moments is the non-negative number $\mathcal{R}({\bm{\xi}},{\bm{\eta}})$ defined by 
        \[
        \mathcal{R}^2({\bm{\xi}},{\bm{\eta}})=\begin{cases}\frac{\mathcal{V}^2({\bm{\xi}},{\bm{\eta}})}{\sqrt{\mathcal{V}^2({\bm{\xi}},{\bm{\xi}})\mathcal{V}^2({\bm{\eta}},{\bm{\eta}})}}&,\ \mathcal{V}^2({\bm{\xi}},{\bm{\xi}})\mathcal{V}^2({\bm{\eta}},{\bm{\eta}})>0\\0&,\ \mathcal{V}^2({\bm{\xi}},{\bm{\xi}})\mathcal{V}^2({\bm{\eta}},{\bm{\eta}})=0\end{cases},\ \text{ where}
        \]
        \[
        \mathcal{V}^{2}({\bm{\xi}},{\bm{\eta}})=\|f_{{\bm{\xi}},{\bm{\eta}}}(t,s)-f_{{\bm{\xi}}}(t)f_{{\bm{\eta}}}(s)\|^{2} =\frac1{c_1c_2}\int_{\mathbb{R}^{p_{\bm{\xi}}+p_{\bm{\eta}}}}\frac{|f_{{\bm{\xi}},{\bm{\eta}}}(t,s)-f_{\bm{\xi}}(t)f_{\bm{\eta}}(s)|^2}{||t||_{2}^{1+p_{\bm{\xi}}}||s||_{2}^{1+p_{\bm{\eta}}}}dtds,
        \]
             $\mathcal{V}^2({\bm{\xi}},{\bm{\xi}})$ and $\mathcal{V}^2({\bm{\eta}},{\bm{\eta}})$ are defined similarly, the non-negative number $\mathcal{V}({\bm{\xi}},{\bm{\eta}})$ is called distance covariance; the positive constants $c_1$ and $c_2$ only depend on the dimensions $p_{\bm{\xi}}$ of ${\bm{\xi}}$ and $p_{\bm{\eta}}$ of ${\bm{\eta}}$, respectively.
        \end{definition}
        The distance correlation can be intuitively thought of as measuring the difference between the characteristic functions of the distribution under the assumption of independence of two random vectors, and the true joint one. Therefore, unlike the Pearson correlation coefficient, the distance correlation is not limited to measuring only linear relationships. 
        
        For finite sample estimation, \cite{szekely2007measuring} presented an alternative definition of empirical distance covariance as follows, which is equivalent to the definition stated in the main text but more convenient for calculations. For data $\{(\bm{\xi}_{(i)},{\bm{\eta}}_{(i)})\}_{i=1}^n$ sampled independently and identically from the joint distribution of $({\bm{\xi}},{\bm{\eta}})$, define
        \[
            \begin{aligned}
                a_{ij}&=||\bm{\xi}_{(i)}-\bm{\xi}_{(j)}||_2,\quad\bar{a}_i.=\frac{1}{n}\sum_{j=1}^na_{ij},\quad\bar{a}_{\cdot_j}=\frac{1}{n}\sum_{i=1}^na_{ij},\\
                \bar{a}_{\cdot\cdot}&=\frac{1}{n^2}\sum_{i,j=1}^na_{ij},\quad A_{ij}=a_{ij}-\bar{a}_{i\cdot}-\bar{a}_{\cdot j}+\bar{a}_{\cdot\cdot},
            \end{aligned}
        \]
            $i,j=1,\dots,n$, ${\bA}=(A_{ij})_{n\times n}$. Similarly, define $b_{ij}=||\bm{\eta}_{(i)}-\bm{\eta}_{(j)}||_2$, $B_{ij}=b_{ij}-\bar{b}_{i\cdot}-\bar{b}_{\cdot j}+\bar{b}_{\cdot\cdot},i,j=1,\dots,n$, $\mathbf{B}=(B_{ij})_{n\times n}$. 
        
        \begin{definition}[Empirical distance covariance]
            The empirical distance covariance $\mathcal{V}_n({\bm{\xi}},{\bm{\eta}})$ is the non-negative number defined by
        \[
                \mathcal{V}_n^2({\bm{\xi}},{\bm{\eta}})=\frac1{n^2}\sum_{i,j=1}^nA_{ij}B_{ij}.
        \]
        \end{definition}
        
        \noindent\textbf{The consistency of $\hat{G}^+$}
        
        \noindent
        \begin{proof}
        Let $P^I_{ij}(n)$ and $P^{II}_{ij}(n)$ denote the probability of making a Type I and Type II error of the DC-based independence test of $X_i$ and $Y_j$, respectively. Firstly, we prove that if we take $\alpha=O(n^{-2})$, then $\lim_{n\to +\infty}P^I_{ij}(n)=\lim_{n\to +\infty}P^{II}_{ij}(n)=0$. Note that $\mathbb{E}(X_i^2)<\infty,\ \mathbb{E}(Y_j^2)<\infty$, for all $i=1,\dots,q,j=1,\dots,p,$ and when $n$ is sufficiently large, $\alpha\le 0.215$. Therefore, by Theorem 6 of \cite{szekely2007measuring}, under the null hypothesis,
        \[
            \lim_{n\to +\infty}P^I_{ij}(n)\le \lim_{n\to +\infty}\alpha= 0.
        \]
        If the alternative hypothesis is true, then by Lemma 2 of \cite{szekely2007measuring}, it follows that
        \[
            T_n/n=\zeta_n=\frac{\mathcal{V}_n^2(X_i,Y_j)}{S_2(X_i,Y_j)}\to_PC>0,
        \]
           where $C$ is a positive constant. For any sufficiently small number $0<\epsilon<C$, we have
        \[
            \lim_{n\to +\infty}\mathbb{P}\{|\zeta_n-C|>\epsilon\}= 0.
        \]
            Therefore, 
        \[
            \begin{aligned}
                P^{II}_{ij}(n)&=\mathbb{P}\bigl\{n\zeta_n\le \big(\Phi^{-1}(1-\alpha/2)\big)^2\bigr\}\\
                &=\mathbb{P}\bigl\{n\zeta_n\le \big(\Phi^{-1}(1-\alpha/2)\big)^2,|\zeta_n-C|>\epsilon\bigr\}+\mathbb{P}\bigl\{n\zeta_n\le \big(\Phi^{-1}(1-\alpha/2)\big)^2,|\zeta_n-C|\le\epsilon\bigr\}\\
                &\le \mathbb{P}\{|\zeta_n-C|>\epsilon\}+\mathbb{P}\biggl\{\zeta_n-C+C\le \big(\Phi^{-1}(1-\alpha/2)\big)^2/n\Bigm| |\zeta_n-C|\le\epsilon\biggr\}\\
                &\le \mathbb{P}\{|\zeta_n-C|>\epsilon\}+\mathbf{1}\bigl\{-\epsilon+C\le \big(\Phi^{-1}(1-\alpha/2)\big)^2/n\bigr\},
            \end{aligned}
        \]
            where $\Phi(\cdot)$ is the standard normal cumulative distribution function. Since $-\epsilon+C>0$, it remains to show that
        \[
            \varlimsup_{n \to \infty}\frac{\Phi^{-1}(1-\alpha/2)}{\sqrt{n}}= 0.
        \]
        Define $x_n=\Phi^{-1}(1-\alpha/2)$, then it is obvious that $x_n\to +\infty$. When $n$ is sufficiently large, we have $1-\Phi(x_n)\le 1/\sqrt{2\pi}\mathrm{exp}(-x_n^2/2)/x_n$ according to the property of the standard normal distribution. Therefore, when $n$ is sufficiently large, 
            \[
            \alpha/2=1-\Phi(x_n)\le \frac{1}{\sqrt{2\pi}x_n}\exp\biggl({-\frac{x_n^2}{2}}\biggr),
            \]
        i.e.,
            \begin{equation}
            \label{eq:normal_tail}
                   \frac{x_n}{\sqrt{n}}\le\frac{2}{\sqrt{2\pi}}\frac{1}{n^{-3/2}}\exp\biggl(-\frac{x_n^2}{2}\biggr).
            \end{equation}
         If $\varlimsup_{n \to \infty}{x_n}/{\sqrt{n}}=\tau>0$, then there exists a subsequence $\{x_{n_i}\}$ such that $\lim_{i\to+\infty}{x_{n_i}}/{\sqrt{n_i}}=\tau>0$. For this subsequence, we have 
            \[
            \lim_{i\to +\infty}\frac{2}{\sqrt{2\pi}n_i^{-3/2}}\exp\biggl(-\frac{x_{n_i}^2}{2}\biggr)=0.
            \]
        This contradicts (\ref{eq:normal_tail}), and thus 
        \[
            \lim_{n\to +\infty}P_{ij}^{II}\le \varlimsup_{n\to +\infty}\mathbb{P}\{|\zeta_n-C|>\epsilon\}+\mathbf{1}\bigl\{-\epsilon+C\le \bigl(\Phi^{-1}(1-\alpha/2)\bigr)^2/n\bigr\}=0.
        \]
        
        Next we prove that $\hat{{G}}^+$ is consistent when $\alpha=O(n^{-2})$. Let $\bR=(R_{ij})_{q\times p}$ where $R_{ij}=\mathbf{1}(X_i\not\! \ind Y_j)$ and $\hat{\bR}$ denote the estimate of $\bR$ as shown in line 2 of Algorithm \ref{alg:peel}. According to Proposition \ref{leafnodes} and Proposition \ref{between}, if $\hat{\bR}=\bR$ then $\hat{{G}}^+={G}^+$, which implies that
        \[
        \mathbb{P}(\hat{{G}}^+\ne{G}^+)\le \mathbb{P}(\hat{\bR}\ne \bR).
        \]
        Moreover,
            \[
            \begin{aligned}
            \mathbb{P}(\hat{\bR}\ne \bR)&=\mathbb{P}\biggl(\bigcup_{i=1}^p\{\hat{\bR}_{i,+}(n)\ne \bR_{i,+}\}\biggr)\le \sum_{i=1}^q\sum_{j=1}^p \bigl\{P^I_{ij}(n)+P^{II}_{ij}(n)\bigr\}.
            \end{aligned}
            \]
        Therefore,
            \[
            \mathbb{P}(\hat{{G}}^+\ne {G}^+)\le \mathbb{P}(\hat{\bR}\ne \bR)\le \sum_{i=1}^q\sum_{j=1}^p \bigl\{P^I_{ij}(n)+P^{II}_{ij}(n)\bigr\}.
            \]
        Since $\lim_{n\to +\infty}P^I_{ij}(n)=\lim_{n\to +\infty}P^{II}_{ij}(n)=0$, it follows that 
            \[
            \lim_{n\to +\infty}\mathbb{P}(\hat{{G}}^+\ne {G}^+)\le \sum_{i=1}^q\sum_{j=1}^p \biggl\{\lim_{n\to +\infty}P^I_{ij}(n)+\lim_{n\to +\infty}P^{II}_{ij}(n)\biggr\}=0.
            \]
        In conclusion, the estimated ARG is consistent, implying that the estimated candidate IV sets are also consistent.
        \end{proof}
        
        \subsection{Proof of Theorem \ref{thm:asympt}}
        
        \begin{proof}
        We first introduce some notations. Let $\mathbf{1}_d$ denote the $d$-dimensional vector with all elements equal to 1. For a vector ${\bv}$, let ${\bv}\odot \mathbf{w}$ the Hadamard product of it with another vector $\mathbf{w}$ of the same dimension. The diagonal matrix with diagonal elements being ${\bv}$ is denoted as $\mathrm{diag}({\bv})$. Let $\bO=(\mathbf{X}_{n\times q},\mathbf{Y}_{n\times p})$ denote the observed data. To distinguish from $\bO_i$ denoting the $i$th feature of a sample $\bO$, we sometimes use $\bO_{(i)}$ to denote the $i$th sample.

        From Theorem \ref{consistency of ARG}, for any Borel set $B\subseteq\mathbb{R}^{|\hat{\mathcal{E}}|^+}$, we have
        \[
            \begin{aligned}
                &\lim_{n\to \infty }\mathbb{P}(\hat{\bbeta}-\bbeta^\circ\in B)\\
                &\quad=\lim_{n\to \infty }\mathbb{P}(\hat{\bbeta}-\bbeta^\circ\in B\mid \hat{{G}}^+={G}^+)\mathbb{P}(\hat{{G}}^+={G}^+)+\mathbb{P}(\hat{\bbeta}-\bbeta^\circ\in B\mid \hat{{G}}^+\ne{G}^+)\mathbb{P}(\hat{{G}}^+\ne{G}^+)\\
                &\quad=\lim_{n\to \infty } \mathbb{P}(\hat{\bbeta}-\bbeta^\circ\in B\mid \hat{{G}}^+={G}^+),
            \end{aligned}
        \]
        where $\bbeta^\circ=(\beta_{kj})_{(k,j)\in \hat{\mathcal{E}}^+}$. It is therefore sufficient to consider the asymptotic distribution of $\hat{\bbeta}$ only when $\hat{G}^+=G^+$, and thus $\bbeta^\circ=\bm{\bbeta^*}$ in such cases. 
        
        We first rewrite equation (\ref{GMM-algorithm}) in Algorithm \ref{alg:gmm} for further discussion. Following the same order of concatenating $\{M_{kj}(\bbeta)\}_{(k,j)\in \mathcal{E}^+}$ to obtain  $\bM(\bbeta)$ denoted by $(k_1,j_1),\dots,(k_{N},j_{N})$, the basis functions $\mathbf{Z}_\gamma(\mathbf{X}_{\ca_G(k)})$ of dimension $t(k;\gamma)$ can be arranged to
        \[
                {\mathbf{Z}}_\gamma:=\left(\mathbf{Z}_\gamma(\mathbf{X}_{\ca_G(k_1)})^T,\ \dots,\ \mathbf{Z}_\gamma(\mathbf{X}_{\ca_G(k_N)})^T\right)^T, 
        \]
        whose dimension is $t_\gamma=\sum_{j=1}^p\sum_{k\in {\an}_G(j)}t(k;\gamma)$. Let the vector $\mathbf{B}_\mathbf{Y} \in \mathbb{R}^{t_\gamma}$ denote $\bigl(Y_{j_1}\mathbf{1}_{t(k_1;\gamma)}^T,\\\dots,Y_{j_{N}}\mathbf{1}_{t(k_N;\gamma)}^T\bigr)^T$. Define $t(k_0;\gamma)=0$ and the matrix ${\bA}_\mathbf{Y}\in \mathbb{R}^{t_\gamma\times N}$ depending on $\mathbf{Y}$ with the $s$th row and $l$th column element given by:
        \[
            ({\bA}_\mathbf{Y})_{sl}=\begin{cases}
            Y_{k_l}& ,\ k_l\in \{k_i\}\cup {\me}_G(k_i,j_i), j_l=j_i \\
            0&  ,\ \text{otherwise}
            \end{cases},\ \text{where }\sum_{\ell=0}^{i-1}t(k_\ell;\gamma)+1\le s\le \sum_{\ell=0}^{i}t(k_\ell;\gamma).
        \]
        In other words, if the $s$th row of $\mathbf{B}_\mathbf{Y}$ corresponds to the part $Y_{j_i}\mathbf{1}^T_{t(k_i;\gamma)}$, then the $s$th row of ${\bA}_\mathbf{Y}\in \mathbb{R}^{t_\gamma\times N}$ satisfies
        \[
            ({\bA}_\mathbf{Y})^T_{s,\cdot}\bbeta=\sum_{l=1}^{N}({\bA}_\mathbf{Y})_{sl}\beta_l=\sum_{\ell\in {\me}_G(k_i,j_i)}\beta_{\ell j_i}Y_\ell+\beta_{k_ij_i}Y_{k_i},
        \]
        for any $\bbeta=(\beta_{k_1j_1},\dots,\beta_{k_Nj_N})^T$. Since ${\mathbf{Z}}_\gamma$ is associated with $\bm{\mu}^*=\mathbb{E}(\mathbf{X})$, let $\hat{\mathbf{Z}}_\gamma$ denote ${\mathbf{Z}}_\gamma$ with $\bm{\mu}^*$ substituted by its estimate $\hat{\bm{\mu}}$. This allows  us to express the equation \eqref{GMM-algorithm} as the following quadratic form equivalently :
        \begin{equation}
        \label{eq:weightedGMM}
            \hat{\bbeta}= \arg\min_{\bbeta} \hat{\mathbb{E}}_n\big\{\hat{\mathbf{Z}}_\gamma\odot (\mathbf{B}_\mathbf{Y}-{\bA}_\mathbf{Y}{\bbeta}) \big\}^T {\bOmega} \hat{\mathbb{E}}_n\big\{\hat{\mathbf{Z}}_\gamma\odot (\mathbf{B}_\mathbf{Y}-{\bA}_\mathbf{Y}{\bbeta}) \big\}.
        \end{equation}
        Therefore,
        \[
            \begin{aligned}
                &\frac{\partial }{\partial  {\bbeta}}\hat{\mathbb{E}}_n\big\{\hat{\mathbf{Z}}_\gamma\odot (\mathbf{B}_\mathbf{Y}-{\bA}_\mathbf{Y} {\bbeta}) \big\}^T {\bOmega} \hat{\mathbb{E}}_n\big\{\hat{\mathbf{Z}}_\gamma\odot (\mathbf{B}_\mathbf{Y}-{\bA}_\mathbf{Y} {\bbeta}) \big\}\Big|_{ {\bbeta}=\hat{\bbeta}}\\
                &\quad = -2\hat{\mathbb{E}}_n(\mathrm{diag}(\hat{\mathbf{Z}}_\gamma){\bA}_{\mathbf{Y}})^T{\bOmega} \hat{\mathbb{E}}_n\big\{\mathrm{diag}(\hat{\mathbf{Z}}_\gamma)(\mathbf{B}_\mathbf{Y}-{\bA}_\mathbf{Y}\hat{\bbeta}) \big\}\\
                &\quad =\mathbf{0}.
            \end{aligned}
        \]
        By concatenating the estimation equations for $\bm{\mu}^*$ and $\bbeta^*$, we have
        \[
            \begin{aligned}
                \begin{bmatrix}
                \hat{\bm{\mu}}\\
                \hat{\bbeta}
                \end{bmatrix}&=\arg\min_{\bm{\mu}, {\bbeta}}\hat{\mathbb{E}}_n\begin{bmatrix}
                \bm{\mu}-\mathbf{X} \\
                \mathrm{diag}(\hat{\mathbf{Z}}_\gamma)(\mathbf{B}_\mathbf{Y}-{\bA}_{\mathbf{Y}} {\bbeta})
                \end{bmatrix}^T\cdot\begin{bmatrix}
                \bI_{q} &\mathbf{0} \\
                \mathbf{0} & {\bOmega}\hat{\mathbb{E}}_n(\mathrm{diag}(\hat{\mathbf{Z}}_\gamma){\bA}_{\mathbf{Y}})\hat{\mathbb{E}}_n(\mathrm{diag}(\hat{\mathbf{Z}}_\gamma){\bA}_{\mathbf{Y}})^T{\bOmega}
                \end{bmatrix}\\
                &\relph{} \cdot \hat{\mathbb{E}}_n\begin{bmatrix}
                \bm{\mu}-\mathbf{X} \\
                \mathrm{diag}(\hat{\mathbf{Z}}_\gamma)(\mathbf{B}_\mathbf{Y}-{\bA}_{\mathbf{Y}} {\bbeta})
                \end{bmatrix}\\
                &=\arg\min_{\bm{\mu}, {\bbeta}}\hat{\mathbb{E}}_n\begin{bmatrix}
                \bm{\mu}-\mathbf{X} \\
                \mathrm{diag}(\hat{\mathbf{Z}}_\gamma)(\mathbf{B}_\mathbf{Y}-{\bA}_{\mathbf{Y}} {\bbeta})
                \end{bmatrix}^T\bW_n\hat{\mathbb{E}}_n\begin{bmatrix}
                \bm{\mu}-\mathbf{X} \\
                \mathrm{diag}(\hat{\mathbf{Z}}_\gamma)(\mathbf{B}_\mathbf{Y}-{\bA}_{\mathbf{Y}} {\bbeta})
                \end{bmatrix},
            \end{aligned}
        \]
        where $\bW_n$ is a data-adaptive positive semi-definite weighting matrix. For ease of notation, define $\hat{\btheta}=(\hat{\bm{\mu}}^T,\hat{\bbeta}^T)^T$, $\btheta^*=(\bm{\mu}^{*T},{\bbeta}^{*T})^T$, and 
        \[
        \bbm_i=\bbm(\bO_{(i)},\btheta):=\begin{bmatrix}
        \bm{\mu}-\mathbf{X}_{(i)} \\
        {\mathrm{diag}(\hat{\mathbf{Z}}_\gamma)}_{(i)}({\mathbf{B}_\mathbf{Y}}_{(i)}-{{\bA}_{\mathbf{Y}}}_{(i)}\bbeta)
        \end{bmatrix},\ for\ i=1,\dots, n.
        \]
        Let $\mathcal{O}\subseteq\mathbb{R}^{p+q}$ denote the sample space of $\bO_{(i)}$, and $\Theta\subseteq\mathbb{R}^{q+ | \mathcal{E}|^+}$ the parameter space of $\btheta$, then $\bbm(\bO_{(i)},\btheta)$ is a mapping from $\mathcal{O}\times \Theta$ to $\mathbb{R}^{q+t_\gamma}$. According to Theorem 3.2 of \cite{hall2005generalized}, we introduce the following regular conditions: 
        \begin{assumption*}[Regularity conditions on $\bbm(\bO_{(i)},\btheta)$]
        \label{regu_1}
            The function $\bbm_i:\mathcal{O}\times \Theta\to \mathbb{R}^{q+t_\gamma}$ satisfies that (i)  it is continuous on $\Theta$ for each $\bO_{(i)}\in \mathcal{O}$; (ii) $\mathbb{E}(\bbm_i)$ is continuous on $\Theta$.
        \end{assumption*}
        
        \begin{assumption*}[Regularity conditions on $\partial \bbm(\bO_{(i)},\btheta)/\partial \btheta^T$]
        \label{regu_2}
            (i) The derivative matrix $\partial \bbm(\bO_{(i)},\btheta)/\partial \btheta^T$ exists and is continuous on $\Theta$ for each $\bO_{(i)}\in \mathcal{O}$; (ii) $\btheta^*$ is an interior point of $\Theta$; (iii) $\mathbb{E}(\partial \bbm(\bO_{(i)},\btheta)/\partial \btheta^T)$exists and is finite.
        \end{assumption*}
        
        \begin{assumption*}[Properties of ${\bOmega}$]
        \label{regu_3}  
        The weighting matrix ${\bOmega}$ may depend on data, but  converges in probability to a positive definite matrix of constants.
        \end{assumption*}
        
        Under Assumption \ref{assumption-L0-norm}, when the positive semi-definite matrix ${\bOmega}$ satisfies the regularity condition \ref{regu_3}, $\bW_n$ convergence in probability to a positive definite matrix $\bW_{\bOmega}$ as $n\to\infty$,
        \[
            \bW_n\to \bW_{\bOmega}=\begin{bmatrix}
            \bI_{q} &\mathbf{0} \\
            \mathbf{0} & {\bOmega} \mathbb{E}(\mathrm{diag}(\mathbf{Z}^*_\gamma){\bA}_{\mathbf{Y}}) \mathbb{E}(\mathrm{diag}(\mathbf{Z}^*_\gamma){\bA}_{\mathbf{Y}})^T{\bOmega}
            \end{bmatrix}.
        \]
        
        \begin{assumption*}[Compactness of $\Theta$]
        \label{regu_4}
            The parameter space $\Theta$ is a compact set.
        \end{assumption*}
        
        \begin{assumption*}[Domination of $\bbm(\bO_{(i)};\btheta)$]
        \label{regu_5}
            $\mathbb{E}\{\sup_{\btheta\in\Theta}\|\bbm(\bO_{(i)},\btheta)\|_2\}<\infty.$
        \end{assumption*}
        
        \begin{assumption*}[Properties of the variance of the sample moment]
        \label{regu_6}
        (i) The moment $\mathbb{E}\{\bbm(\bO_{(i)},\btheta^*)\bbm(\bO_{(i)},\btheta^*)^T\}$ exists and is finite; (ii) The limit 
        \[
            \lim_{n\to+\infty}\ \mathrm{var}\biggl(\sqrt{n}\hat{\mathbb{E}}_n\begin{bmatrix} \bm{\mu}^*-\mathbf{X} \\
            \mathrm{diag}(\mathbf{Z}^*_\gamma)(\mathbf{B}_\mathbf{Y}-{\bA}_{\mathbf{Y}}{\bbeta}^*)
            \end{bmatrix}\biggr)
        \] 
        exists and is finite.
        \end{assumption*}
        
        \begin{assumption*}[Continuity of $\mathbb{E}\bigl\{\partial \bbm(\bO_{(i)},\btheta)/\partial \btheta^T\bigr\}$]
        \label{regu_7}
            The function $\mathbb{E}\bigl\{\partial \bbm(\bO_{(i)},\btheta)/\partial \btheta^T\bigr\}$ is continuous on some neighborhood $N_\epsilon$ of $\btheta^*$.
        \end{assumption*}
        
        \begin{assumption*}[Uniform Convergence of $\bG_n(\btheta)=n^{-1}\sum_{i=1}^n\partial \bbm(\bO_{(i)},\btheta)/\partial\btheta^T$]
        \label{regu_8}
        \[
            \operatorname{sup}_{\btheta\in N_{\epsilon}}\bigl\|\bG_{n}(\btheta)-\mathbb{E}\bigl\{\partial \bbm(\bO_{(i)},\btheta)/\partial\btheta^{T}\bigr\}\bigr\|_2\to_P0.
        \]
        \end{assumption*}
        
        Suppose that Assumptions \ref{assumption-independent X}--\ref{assumption-L0-norm} and the regularity conditions \ref{regu_1}--\ref{regu_7} hold. According to Theorem 3.2 of \cite{hall2005generalized}, we have
        \[
            \sqrt{n}(\hat{\btheta}-\btheta^*)\to_d N\left(\mathbf{0},(\bG^T\bW_{\bOmega} \bG)^{-1}\bG^T\bW_{\bOmega} \bF \bW_{\bOmega}  \bG(\bG^T\bW_{\bOmega} \bG)^{-1}\right),
        \]
        where 
        \[
            \begin{aligned}
                    \bG&=\mathbb{E}\Biggl\{\frac{\partial \bbm(\bO_{(i)},\btheta)}{\partial \btheta}\Bigg|_{\btheta=\btheta^*}\Bigg\} =\begin{bmatrix}
                \bI_{q}  &\mathbf{0} \\
                \bC  & \bD
                \end{bmatrix},\\
                \bF&=\lim_{n\to+\infty}\ \text{var}\bigg\{\sqrt{n}\hat{\mathbb{E}}_n\begin{bmatrix}
                \bm{\mu}^*-\mathbf{X} \\
                \mathrm{diag}(\mathbf{Z}^*_\gamma)(\mathbf{B}_\mathbf{Y}-{\bA}_{\mathbf{Y}}{\bbeta}^*)
                \end{bmatrix}\bigg\},
            \end{aligned}
        \]
        with 
        \[
            \begin{aligned}
                \bC&=\mathbb{E}\bigg\{\frac{\partial }{\partial \bm{\mu}}\mathrm{diag}(\hat{\mathbf{Z}}_\gamma)(\mathbf{B}_\mathbf{Y}-{\bA}_{\mathbf{Y}} {\bbeta})\bigg|_{\bm{\mu}=\bm{\mu}^*, {\bbeta}={\bbeta}^*} \bigg\},\\
                \bD&=\mathbb{E}\bigg\{\frac{\partial }{\partial  {\bbeta}}\mathrm{diag}(\hat{\mathbf{Z}}_\gamma)(\mathbf{B}_\mathbf{Y}-{\bA}_{\mathbf{Y}} {\bbeta})\bigg|_{\bm{\mu}=\bm{\mu}^*, {\bbeta}={\bbeta}^*}\bigg\}=-\mathbb{E}\{\mathrm{diag}(\mathbf{Z}^*_\gamma){\bA}_{\mathbf{Y}}\}.
            \end{aligned}
        \]
        In particular, the matrix $\bC$ depends on the form of $\hatbasisk$. For example, when the candidate IVs all take values in $\{0,1\}$,  
        \[
            \hatbasisk=\bigl(\Pi_{s\in \alpha_k(1)}(X_s-\hat{\mu}_s),\dots,\Pi_{s\in \alpha_k(t(k;\gamma))}(X_s-\hat{\mu}_s)\bigr)^T,
        \]
        where $\alpha_k(\cdot)$ are the elements of $\{\alpha :\alpha\subseteq \ca_G(k),\ |\alpha|\ge |\ca_G (k)|-\gamma+1\}$ as described in the main text. Therefore for each $j=1,\dots,q$, the partial derivation is 
        \[
            \frac{\partial \hatbasisk}{\partial \mu_j}\Bigg|_{\bm{\mu}=\bm{\mu}^*,\bbeta=\bbeta^*} =\biggl[\biggl\{-\mathbf{1}(j\in\alpha_k(\ell))\prod_{s\in\alpha_k(\ell),s\ne j}(X_s-\mu^*_s)\biggr\}_{\ell=1}^{t(k;\gamma)}\biggr]^T.
        \]
        Since we are only interested in the asymptotic properties of  $\hat{\bbeta}$, Theorem \ref{thm:asympt} holds by extracting the corresponding submatrix. 
        \end{proof}
        
        \subsection{Proof of Theorem \ref{thm:fdr}}
        
        \begin{proof}
        Under Assumptions \ref{assumption-independent X}--\ref{assumption-valid IV}, it has been shown that $\hat{{G}}^+$ is consistent according to Theorem \ref{consistency of ARG}. Moreover, as long as $\hat{{G}}^+=G^+$, we have $\RE(n)=0$. Therefore,
        \[
            \begin{aligned}
                &\lim_{n\to \infty}\mathbb{E}\{\RE(n)\}\\
                &\quad=\lim_{n\to \infty}\mathbb{E}\{\RE(n) \mid \hat{{G}}^+\ne {G}^+\}\mathbb{P}(\hat{{G}}^+\ne {G}^+\big)+\lim_{n\to \infty}\mathbb{E}\{\RE(n) \mid \hat{{G}}^+= {G}^+\}\mathbb{P}(\hat{{G}}^+= {G}^+)\\
                &\quad\le  | \mathcal{E}| \lim_{n\to \infty} \mathbb{P}(\hat{{G}}^+\ne {G}^+)+0\\
                &\quad =0.
            \end{aligned}
        \]
        For any given $0<q^*<1$, we take it as the target control level of FDR $q^*$ in Algorithm \ref{alg:gmm}. Then according to Theorem 1.3 of \cite{benjamini2001control}, we have
        \[
            \lim_{n\to \infty}\mathbb{E}\Biggl\{\frac{\FP(n)}{\TP(n)+\RE(n)+\FP(n)}\Biggm| \hat{{G}}^+={G}^+\Biggr\}\le q^*.
        \]
        Since the $\hat{{G}}^+$ is consistent and $\mathcal{E}\ne \emptyset$, it follows that
        \[
            \begin{aligned}
                &\lim_{n\to\infty}\FDR(\hat{\mathcal{E}})\\
                &\quad=\lim_{n\to\infty}\mathbb{E}\biggl\{\frac{\RE(n)}{\TP(n)+\RE(n)+\FP(n)}\biggr\}+\lim_{n\to\infty}\mathbb{E}\biggl\{\frac{\FP(n)}{\TP(n)+\RE(n)+\FP(n)}\biggr\}\\
                &\quad=0+\lim_{n\to\infty}\mathbb{E}\biggl\{\frac{\FP(n)}{\TP(n)+\RE(n)+\FP(n)}\biggm|\hat{{G}}^+={G}^+\biggr\}\mathbb{P}(\hat{{G}}^+={G}^+)\\
                &\quad\relph{}+\lim_{n\to\infty}\mathbb{E}\biggl\{\frac{\FP(n)}{\TP(n)+\RE(n)+\FP(n)}\biggm|\hat{{G}}^+\ne{G}^+\biggr\}\mathbb{P}(\hat{{G}}^+\ne{G}^+)\\
                &\quad =\lim_{n\to\infty}\mathbb{E}\biggl\{\frac{\FP(n)}{\TP(n)+\RE(n)+\FP(n)}\biggm|\hat{{G}}^+={G}^+\biggr\}\\
                &\quad \le q^*,
            \end{aligned}
        \]
        which completes the proof. 
        \end{proof}
        
        \section{Partial residual plots for ADNI data}\label{sec:partial}
\begin{figure}
            \centering
            \includegraphics[width=0.96\linewidth]{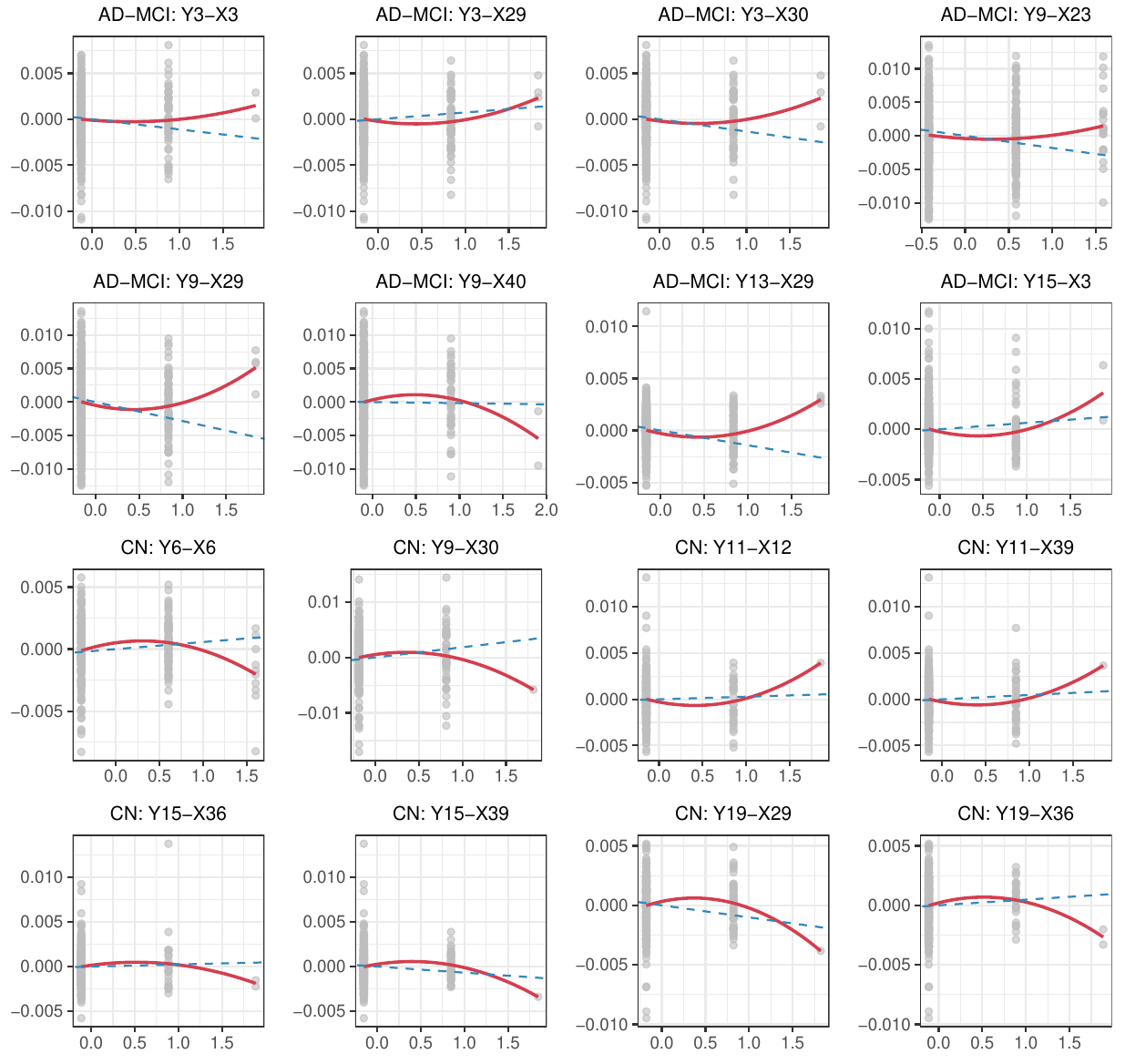}
            \caption{Partial residual plots for the AD-MCI and CN groups.}
            \label{fig:prp}
        \end{figure}

        In this appendix, we present partial residual plots to visualize the nonlinear relationships between certain primary and secondary variables in the ADNI data. Partial residual plots are a useful graphical diagnostic for detecting curvature in regression analysis \citep{cook1993exploring}. Figure \ref{fig:prp} shows the partial residual plots for regression of $Y_j$ on $X_\ell$ adjusting for the identified causal effects and all other secondary variables, where a deviation between the fitted curve and the regression line suggests a nonlinear relationship. From these plots we see that some primary and secondary variables exhibit substantially nonlinear relationships.
        
	\bibliographystyle{apalike}
\bibliography{placid}

\end{document}